\documentclass[fleqn,usenatbib]{mnras}
\usepackage{newtxtext,newtxmath}
\usepackage[T1]{fontenc}
\usepackage{graphicx}
\usepackage{amsmath}
\usepackage{multicol}
\usepackage{amsfonts}
\usepackage{booktabs}
\usepackage{newtxtext,newtxmath}
\usepackage{nicematrix}
\usepackage{caption}
\usepackage{enumitem}
\usepackage[utf8]{inputenc}
\usepackage{makecell}
\usepackage{threeparttable}
\usepackage{tabularray}
\usepackage{cleveref}
\usepackage{array}
\usepackage{float}

\title[Dynamical influence on the evolution of GC pulsars]{Influences of dynamical disruptions on the evolution of pulsars in globular clusters}

\author[K. Oh et al.]{
	Kwangmin Oh,$^{1}$
	C. Y. Hui,$^{2}$\thanks{Corresponding author E-mail: huichungyue@gmail.com}
	Jongsuk Hong,$^{3}$\thanks{Corresponding author E-mail: jshong@kasi.re.kr}
	J. Takata,$^{4}$
	A. K. H. Kong,$^{5}$
	Pak-Hin Thomas Tam,$^{6}$
	\newauthor
	Kwan-Lok Li,$^{7}$
	~and~K. S. Cheng$^{8}$
	\\
	$^{1}$Department of Space Science and Geology, Chungnam National University, Daejeon 34134, Korea\\
	$^{2}$Department of Astronomy and Space Science, Chungnam National University, Daejeon 34134, Korea\\
	$^{3}$Korea Astronomy and Space Science Institute, Daejeon 34055, Republic of Korea\\
	$^{4}$ Department of Astronomy, School of Physics, Huazhong University of Science and Technology, Wuhan 430074, People's Republic of China \\
	$^{5}$ Institute of Astronomy, National Tsing Hua University, Hsinchu 30013, Taiwan\\
	$^{6}$ School of Physics and Astronomy, Sun Yat-sen University, Guangzhou 510275, People's Republic of China\\
	$^{7}$ Department of Physics, National Cheng Kung University, 701401 Tainan, Taiwan\\
	$^{8}$ Department of Physics, The University of Hong Kong, Pokfulam Road, 999077, Hong Kong
}

\begin{document}
	\label{firstpage}
	\pagerange{\pageref{firstpage}--\pageref{lastpage}}
	\maketitle

	\begin{abstract}
		By comparing the physical properties of pulsars hosted by core-collapsed (CCed) and non-core-collapsed (Non-CCed) globular clusters (GCs), we find that pulsars in CCed GCs rotate significantly slower than their counterparts in Non-CCed GCs. Additionally, radio luminosities at 1.4~GHz in CCed GCs are higher. These findings are consistent with the scenario that dynamical interactions in GCs can interrupt angular momentum transfer processes and surface magnetic field decay during the recycling phase. Our results suggest that such effects in CCed GCs are stronger due to more frequent disruptions of compact binaries. This is further supported by the observation that both estimated disruption rates and the fraction of isolated pulsars are predominantly higher in CCed GCs.
	\end{abstract}

	
	\section{Introduction}
	
	Millisecond pulsars (MSPs) are characterized by fast rotations with rotational periods $P_{\rm rot}$ typically shorter than a few tens of milliseconds and relatively weak surface magnetic fields $B_{s}\lesssim10^{9}$~G \citep{Manchester2005.129,2019Galax...7...93H}.
	In order to achieve such fast rotation, MSPs are generally believed to have gone through an accretion phase, during which neutron stars gain angular momentum transferred from their companion stars \citep{Alpar1982.300,Radhakrishnan...Srinivasan1982.51,Fabian1983.301}. This is commonly referred to as the recycling process. During recycling, mass accretion on the neutron star surface can potentially lead to magnetic field decay, as shown in \citep{Cumming2004ApJ.609.999C}, which might account for the weak dipolar field strength inferred from observations.
	
	MSPs can be further separated into two subgroups according to their locations: those residing in globular clusters (GCs) and those in the Galactic field (GF). Owing to the high stellar densities in GCs, the formation of MSPs inside a cluster can be influenced by intracluster dynamical processes \citep[cf.][]{Sigurdsson_1995,Ivanova_2008,Hui2010.714,Ye_2019}. While primary encounter interactions, such as tidal capture or direct collision with a giant, can facilitate binary formation \citep{Fabian_1975, Press_1977, Lee_1986, Lombardi_2006, Fregeau_2007, Ye_2022}, subsequent encounters (referred to as secondary encounters hereafter) can play a role in disrupting binaries \citep{Verbunt2014.561}.
	
	Many studies have shown that dynamical interactions in GCs can lead to an increase in MSP population in comparison with the GF MSPs which rely on binary evolution alone \citep[e.g.][]{Ye_2019,Hui2010.714,Ivanova_2008}. This is consistent with the well-known fact that the formation rate per unit mass of low-mass X-ray binaries (LMXBs), which are the progenitors of MSPs, is orders of magnitude larger in GCs than in GF \citep{1975Natur.253..698K,1975ApJ...199L.143C}. Although many more LMXBs can be assembled in GCs, the mass-transferring processes can be interrupted by the subsequent encounters \citep{verbunt2014}. Such intricate dynamics could potentially lead to differences in the properties of MSPs in GCs compared to those in GF.
	
	The sample sizes of the currently known populations of MSPs in GF and GCs are comparable, which allows a reasonable comparison of the properties between these two populations. In a recent study, \cite{Lee_2023} performed a systematic comparison of rotational, orbital, and X-ray properties of MSPs in GCs and GF. They found that MSPs in GCs generally rotate slower than those in the GF. There is also an indication that the surface magnetic field of GC MSPs is stronger than those in the GF. These findings are consistent with the scenario that the recycling processes of GC MSPs were interrupted by secondary encounters, leading to shortened epochs for both angular momentum transfer and possible magnetic field decay.
	
	Based on the photometric concentrations, GCs can be classified into core-collapsed (CCed) and non-core-collapsed (Non-CCed) \cite[][2010 edition]{Harris1996.112}. A core collapse in a GC is likely a result of gravothermal instability \citep[cf.][]{Lynden-Bell_1968}, which can significantly affect the kinematic properties. 
	
	While the number of X-ray sources in GCs generally correlates with the primary encounter rate $\Gamma$\footnote{$\Gamma\propto\rho_{c}^{1.5}r_{c}^{2}$, where $\rho_{c}$ and $r_{c}$ are the density and radius of the cluster core respectively.} \citep{Pooley_2003}, \cite{Bahramian_2013} found that CCed GCs have fewer X-ray sources than Non-CCed GCs for the same value of $\Gamma$ (see Figure 9 in \citet{Bahramian_2013}). This might indicate the dynamical status of CCed GCs is different from that of Non-CCed GCs, which can leave an imprint on the evolution of compact binaries. Therefore, it is reasonable to speculate that the properties of GC MSPs may be further diversified between CCed and Non-CCed GCs. 
	
	Motivated by the aforementioned findings, we aim to explore potential differences in the properties of pulsars within CCed and Non-CCed GCs by conducting a statistical analysis of selected parameters. In Section 2, we describe our procedure for preparing the data for analysis. The results of statistical analysis are given in Section 3 and their implications will be discussed in Section 4.  
	
	\begin{figure*}
		\begin{multicols}{2}
			\includegraphics[width=\linewidth]{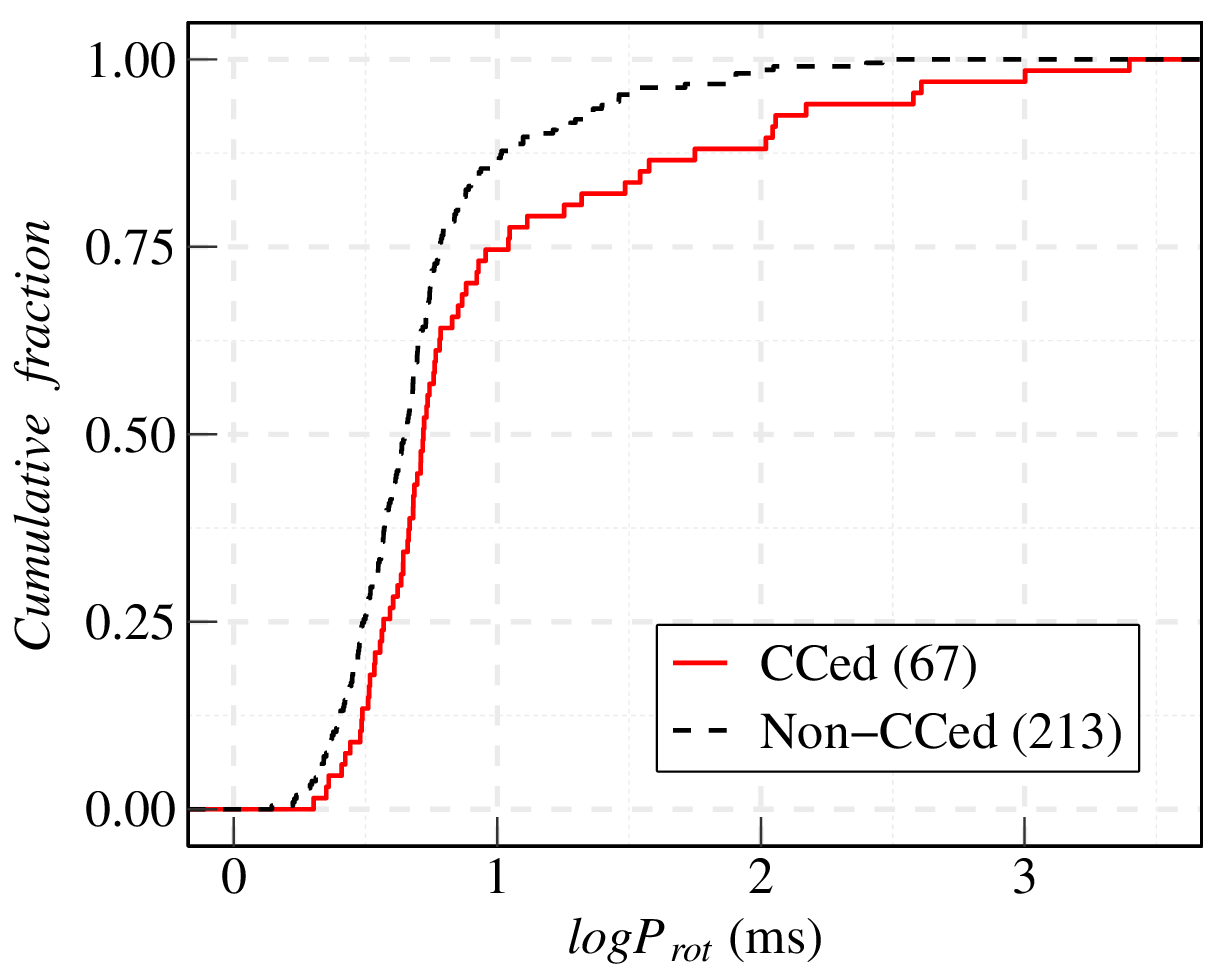}\par
			\includegraphics[width=\linewidth]{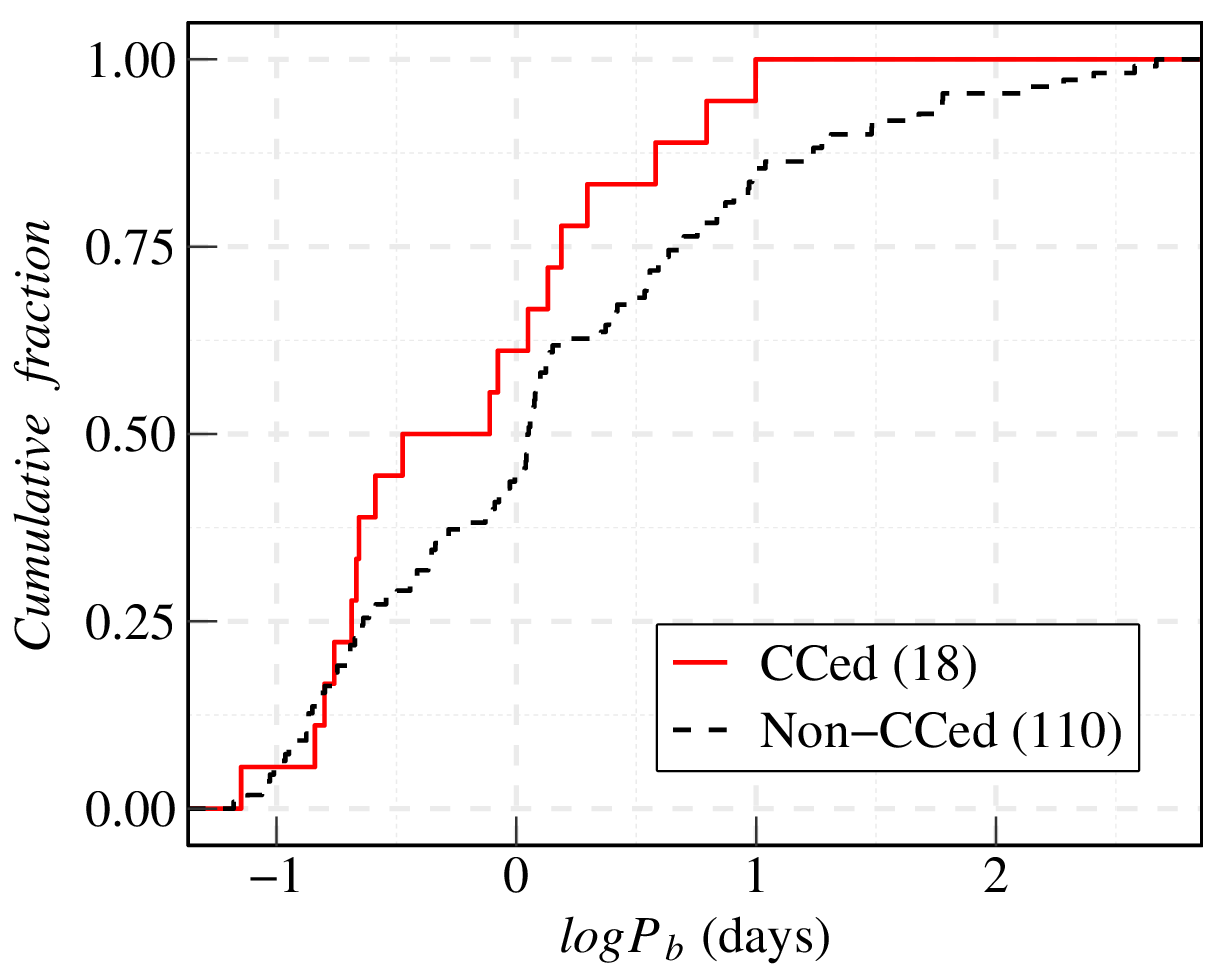}\par
		\end{multicols}
		\begin{multicols}{2}
			\includegraphics[width=\linewidth]{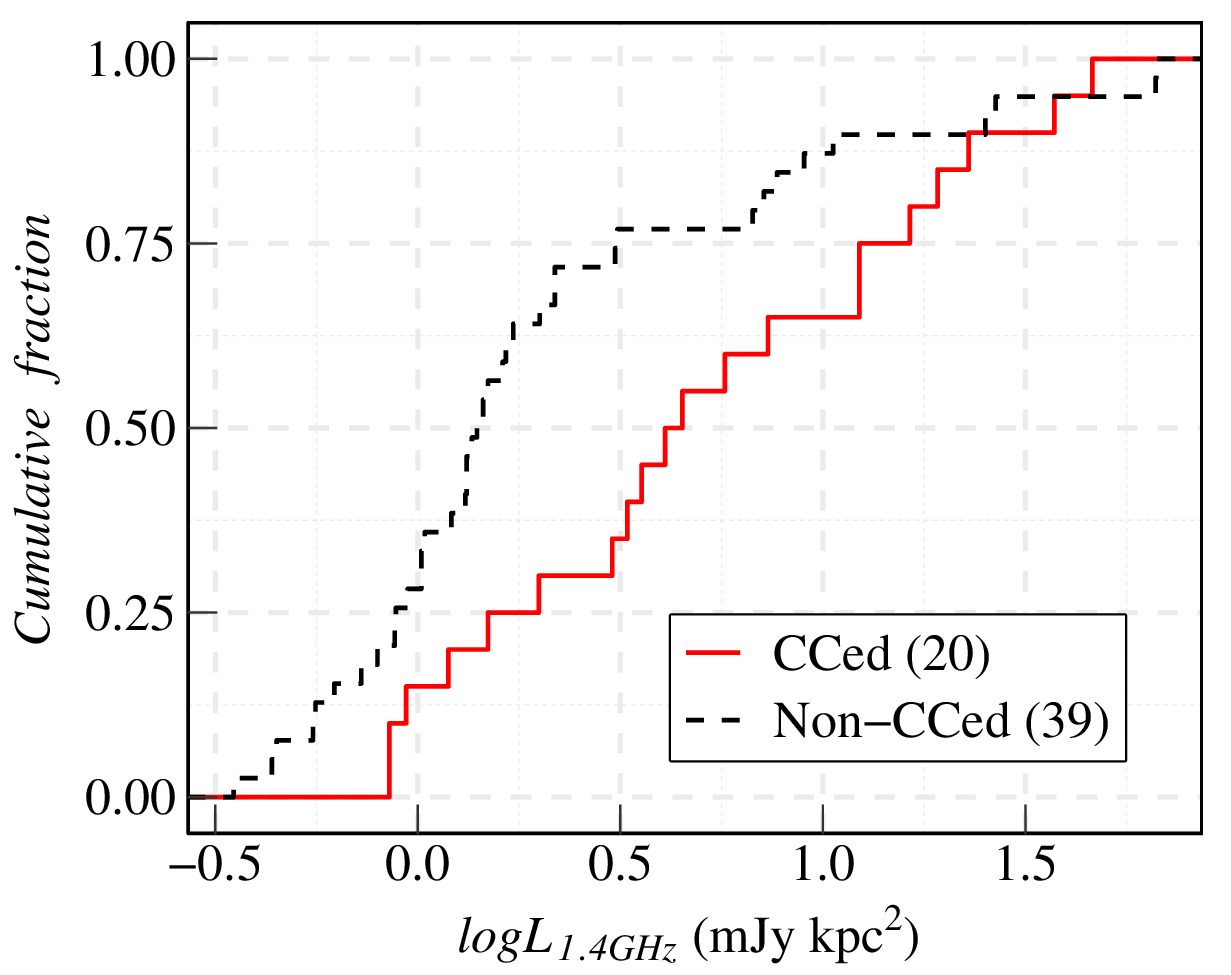}\par
			\includegraphics[width=\linewidth]{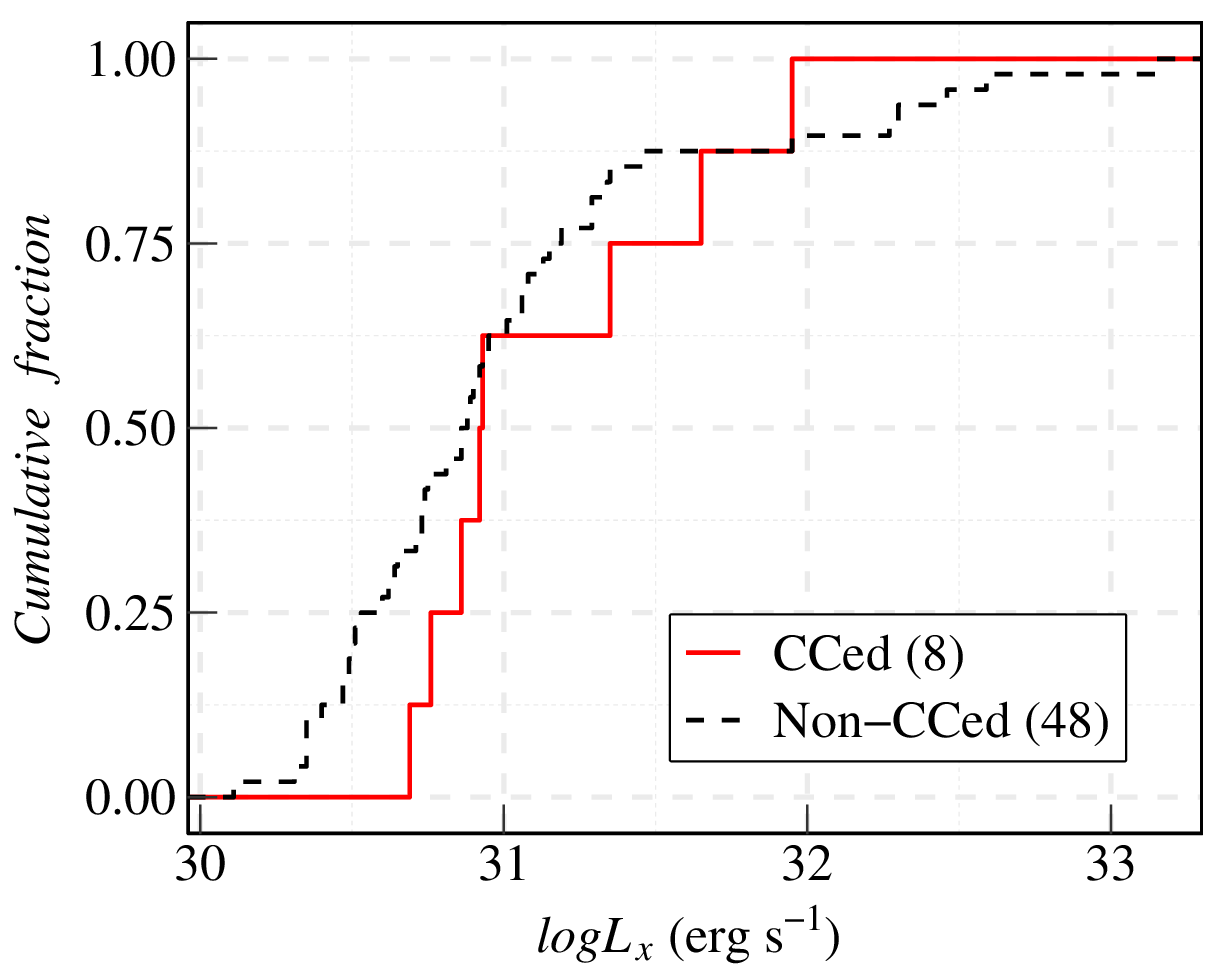}\par
		\end{multicols}
		\vspace{-0.5cm}
		\caption{Comparisons of eCDFs of the selected pulsar properties between CCed GCs and Non-CCed GCs. The bracketed numbers in the legends show the corresponding sample sizes.}
		\label{fig:cdf}    
	\end{figure*}
	
	\begin{figure*}
		\begin{multicols}{2}
			\includegraphics[width=\linewidth]{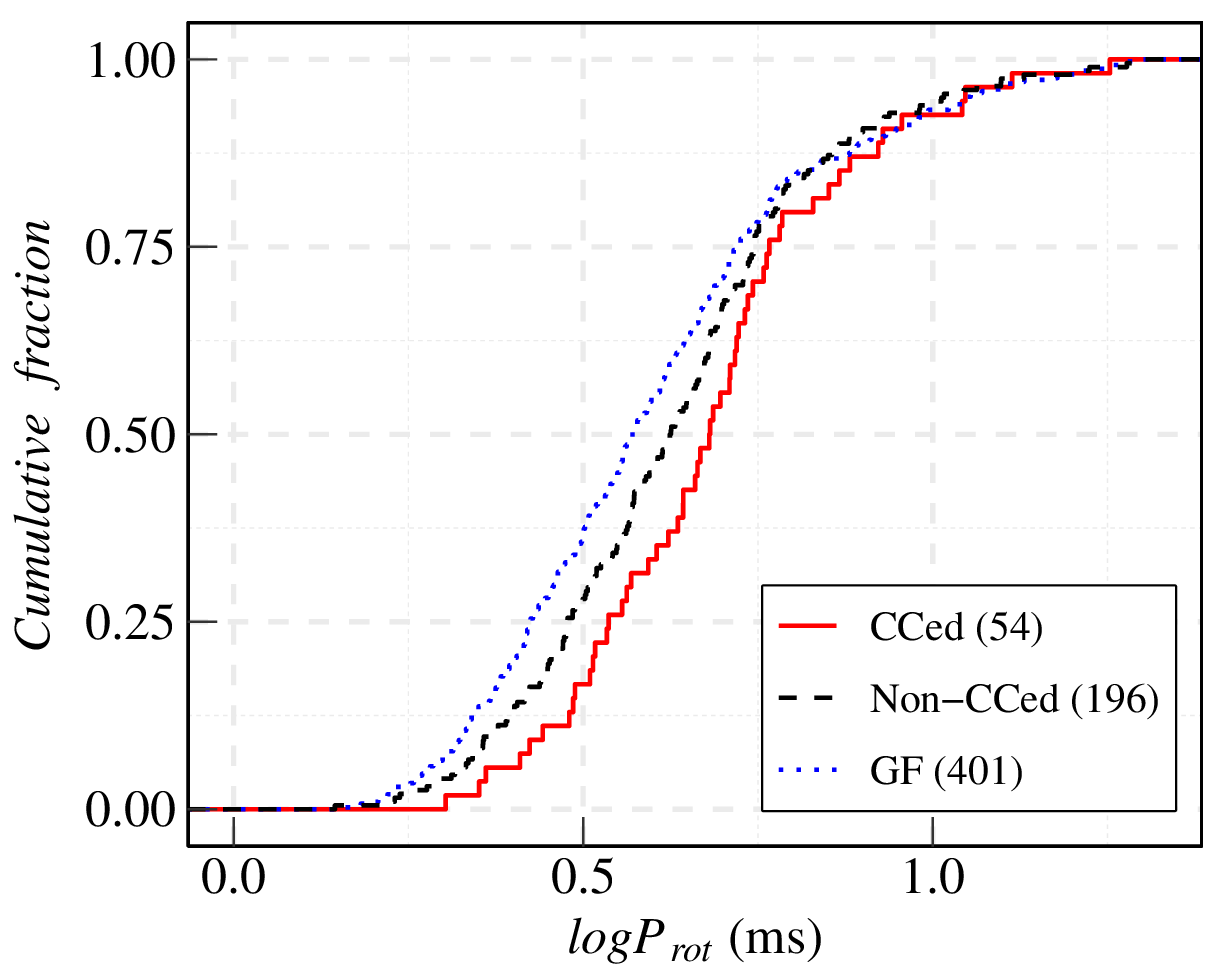}\par
			\includegraphics[width=\linewidth]{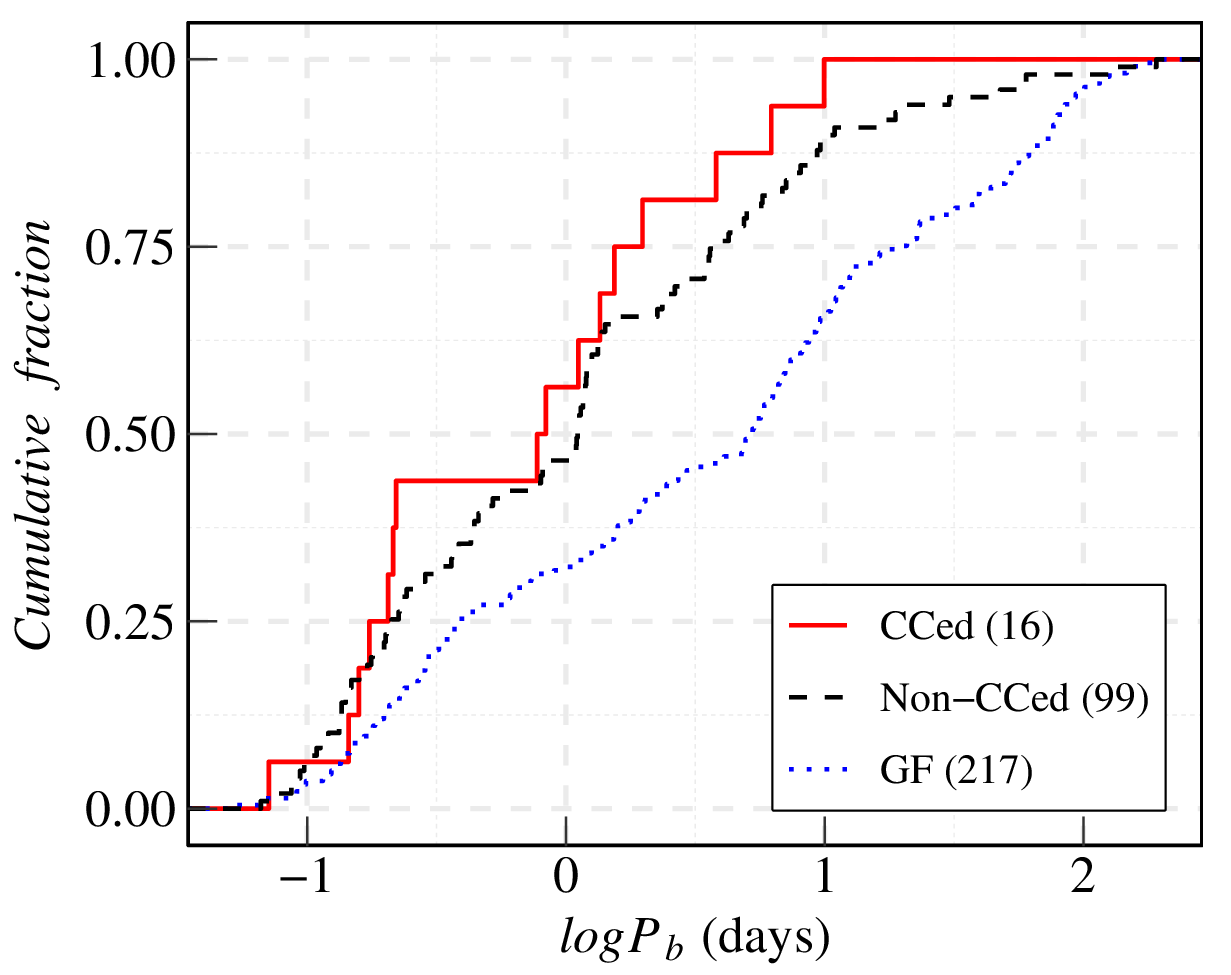}\par
		\end{multicols}
		\begin{multicols}{2}
			\includegraphics[width=\linewidth]{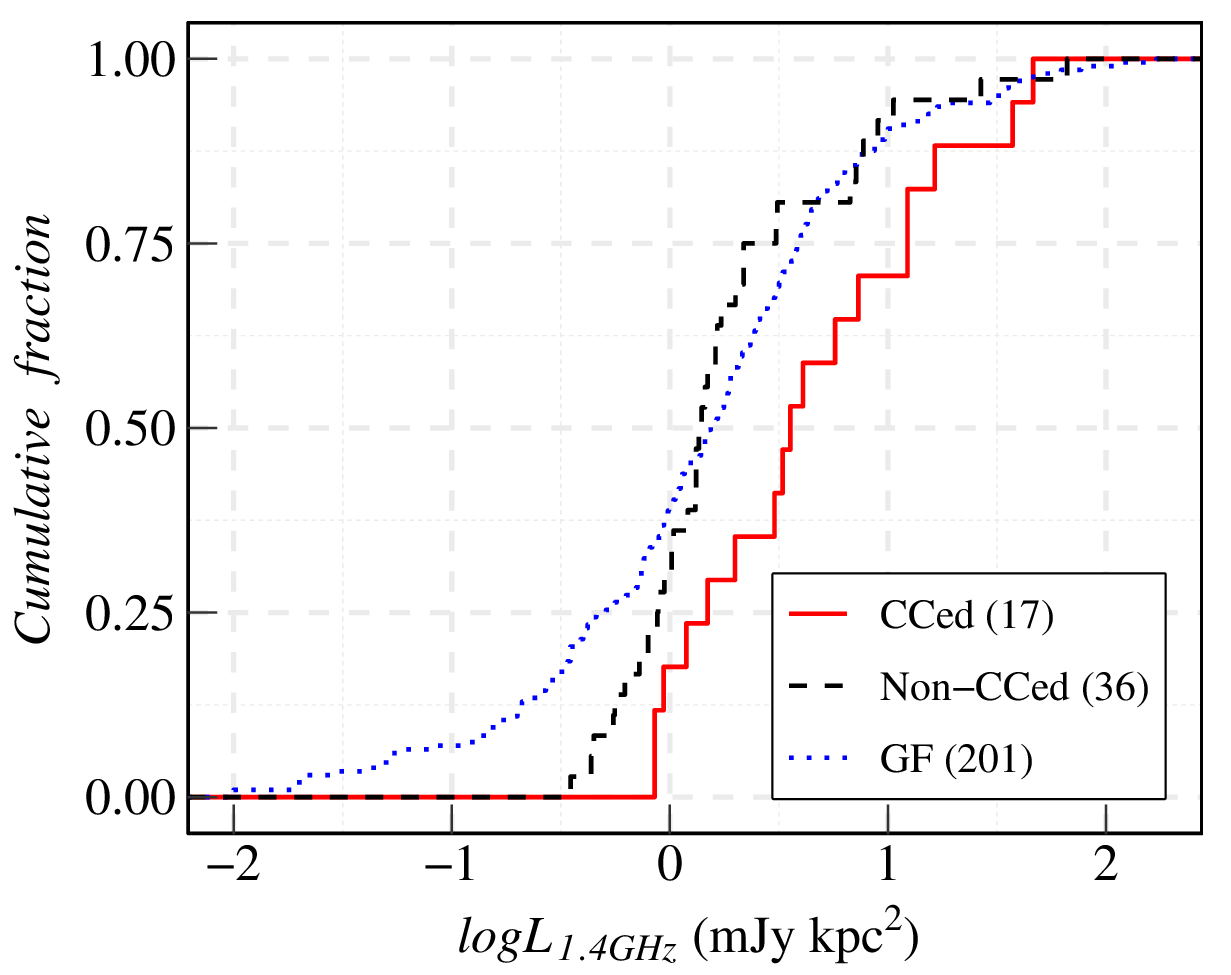}\par
			\includegraphics[width=\linewidth]{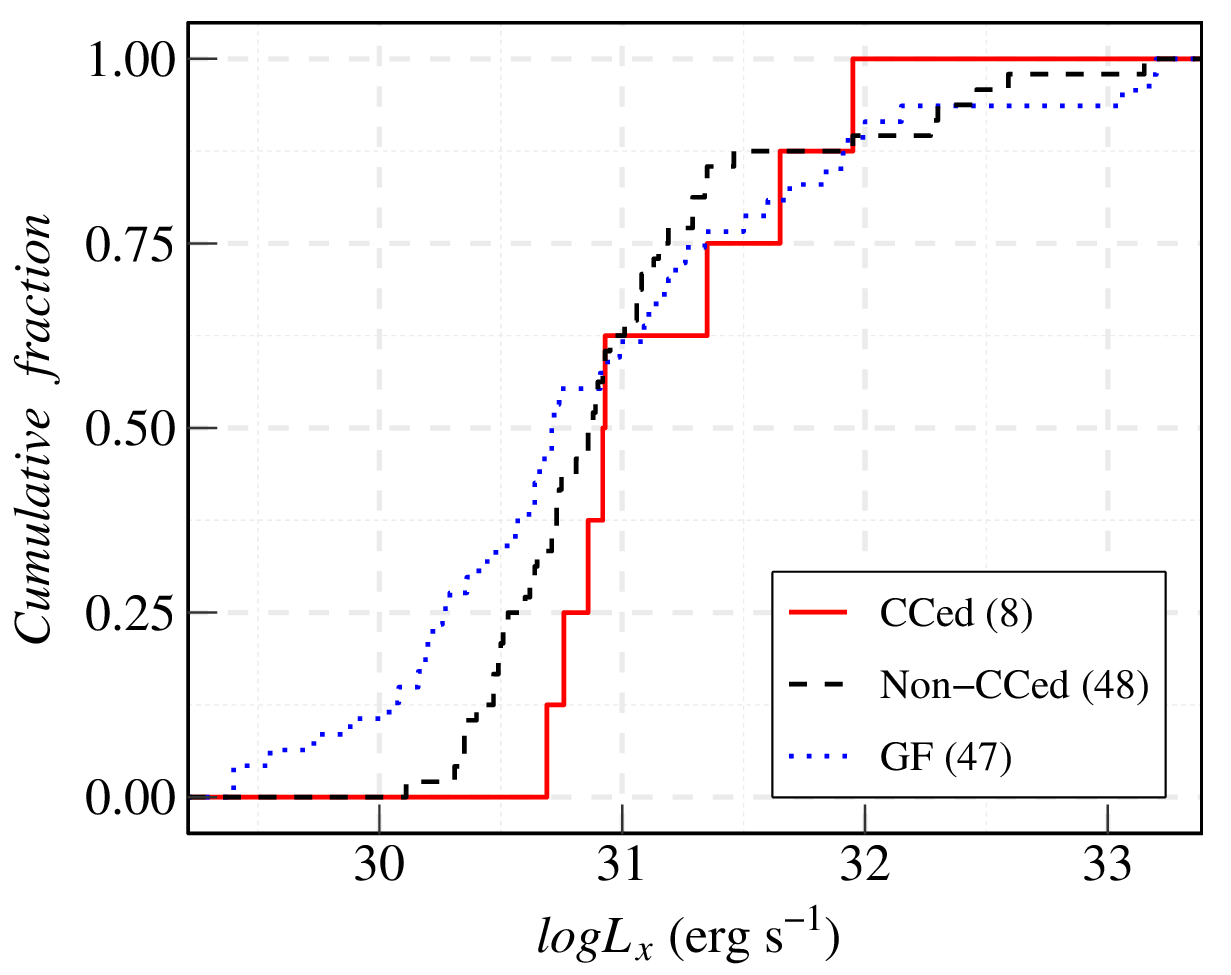}\par
		\end{multicols}
		\vspace{-0.5cm}
		\caption{Comparisons of eCDFs of the selected pulsar properties among CCed GCs, Non-CCed GCs, and GF. The bracketed numbers in the legends show the corresponding sample sizes.}
		\label{fig:cdf2}    
	\end{figure*}
	
	\section{Data Preparation}
	First, we have selected a sample of 280 pulsars from 38 different GCs from the Australia Telescope National Facility (ATNF) Pulsar Catalogue \citep[][ver. 1.70]{Manchester2005.129}. In this work, we only collected the following parameters from the Catalogue: rotational period $P_{\rm rot}$, orbital period $P_{b}$, radio luminosity in L-band $L_{\rm 1.4GHz}$. On the other hand, we have adopted the X-ray luminosities $L_{x}$ (0.3-8~keV) of 56 X-ray emitting MSPs from Table~2 in \cite{Lee_2023}. 
	
	Observationally, it is a common practice to classify whether a GC is CCed or Non-CCed based on its surface brightness profile \citep[e.g.][]{1995AJ....109..218T, Harris1996.112, 2018MNRAS.475.4841R}. Owing to the increased stellar density towards the cluster centre, a GC is defined as CCed if its surface brightness profile  exhibits a power law until the limit of observational resolution \citep{2018MNRAS.475.4841R,1995AJ....109..218T}. On the other hand, Non-CCed GCs typically exhibit a flattened profile towards their centres and follow a King profile \citep{1995AJ....109..218T,1966AJ.....71...64K}. 
	
	In Section 3.1, we adopted the classifications given by \citet[][2010 version]{Harris1996.112} in determining whether a GC is CCed or Non-CCed. Using these labels, we divided our samples accordingly and compared their properties.

	\section{Statistical Analysis \& Results}
	\subsection{Core-Collapsed GCs versus Non-Core-Collapsed GCs}
	We conducted a detailed statistical analysis to compare the aforementioned selected properties of pulsars in CCed and Non-CCed GCs. For each population, we first constructed the unbinned empirical cumulative distribution function (eCDF) of the parameters which are shown in \autoref{fig:cdf}. By visual inspection, these properties appear to be different between these two populations. To quantify the possible difference, we used a two-sample Anderson-Darling (A-D) test to investigate whether such differences are significant. In this work, we consider the difference between two eCDFs to be significant if the $p$-values inferred from A-D test is $<0.05$. The results of the A-D test are summarized in \autoref{tab:adtest}.
	
	We found that the distributions of $P_{\rm rot}$ and $L_{\rm 1.4GHz}$ from CCed GCs and Non-CCed GCs are significantly different. The corresponding $p-$values inferred from A-D test are found to be $0.003$ and $0.014$ respectively. From the distributions of $P_{\rm rot}$ as shown in the upper-right panel of \autoref{fig:cdf}, one can see that the pulsars in CCed GCs  generally rotate slower than those in Non-CCed GCs. The median of $P_{\rm rot}$ in CCed and Non-CCed populations are $5.24$~ms and $4.45$~ms respectively.
	
	For $L_{\rm 1.4GHz}$, the distributions of these two groups of GC pulsars are obviously different (lower-left panel of \autoref{fig:cdf}). It is very clear that the pulsars in the CCed GCs are more powerful radio emitters. The median of $L_{\rm 1.4GHz}$ in CCed and Non-CCed are found to be $4.29$~mJy~kpc$^{2}$ and $1.4$~mJy~kpc$^{2}$ respectively. 
	
	\autoref{fig:cdf} suggests that $P_{b}$ of the pulsars in CCed GCs are shorter, indicating that they have tighter orbits compared to those in Non-CCed GCs. This finding is consistent with our understanding that pulsars with longer orbital periods in CCed GCs are more likely to have been disrupted by dynamical interactions. However, the $p$-value obtained from the A-D test is 0.24, which falls short of our pre-defined criterion for claiming a significant difference between the two groups. This result may be due to the small sample size.
	
	While \cite{Lee_2023} have compared the MSP properties between the GF and GC populations, and identified differences between these them, they did not separately compare GF MSPs with those in CCed GCs and Non-CCed GCs. To complement the analysis conducted by \cite{Lee_2023} as well as our aforementioned investigations, we have further compared MSP properties among the populations in the GF, CCed GCs, and Non-CCed GCs.
	
	In comparing with the MSP properties in the GF, we have followed the same selection criterion as in \cite{Lee_2023}, by selecting pulsars with $P_{\rm rot}<20$~ms in all three populations (i.e. GF, CCed GCs and Non-CCed GCs). This procedure can avoid including non-recycled GF pulsars in this part of the analysis. The eCDFs of $P_{\rm rot}$, $P_{b}$, $L_{\rm 1.4GHz}$, and $L_{x}$ are shown in \autoref{fig:cdf2}. The results of the A-D tests are summarized in \autoref{tab:adtest}.
	
	From the distribution of $P_{\rm rot}$, it is obvious that the rotation of MSPs in the GF is significantly faster than those in CCed and Non-CCed GCs. Moreover, we can see that the difference between CCed GCs and GF is larger than that between Non-CCed GCs and GF. We also find that the $P_{b}$ of GF MSPs is significantly longer than those in GCs, regardless of whether they are CCed or Non-CCed. All these findings align with the scenario suggested by \cite{Lee_2023}, which posits that intracluster dynamics have resulted in the formation of more tightly-bound binaries and the interruptions of the recycling process.
	
	For comparing the distributions of luminosities between GF and GC MSPs in X-ray and radio, we found differences that are statistically acceptable (see \autoref{tab:adtest}). However, given the current sample, it is difficult to rule out the possibility that such differences have resulted from the observational bias between GF and GCs (see the discussion in Section 4).  
	
\begin{figure*}
		\begin{multicols}{2}
			\includegraphics[width=\linewidth]{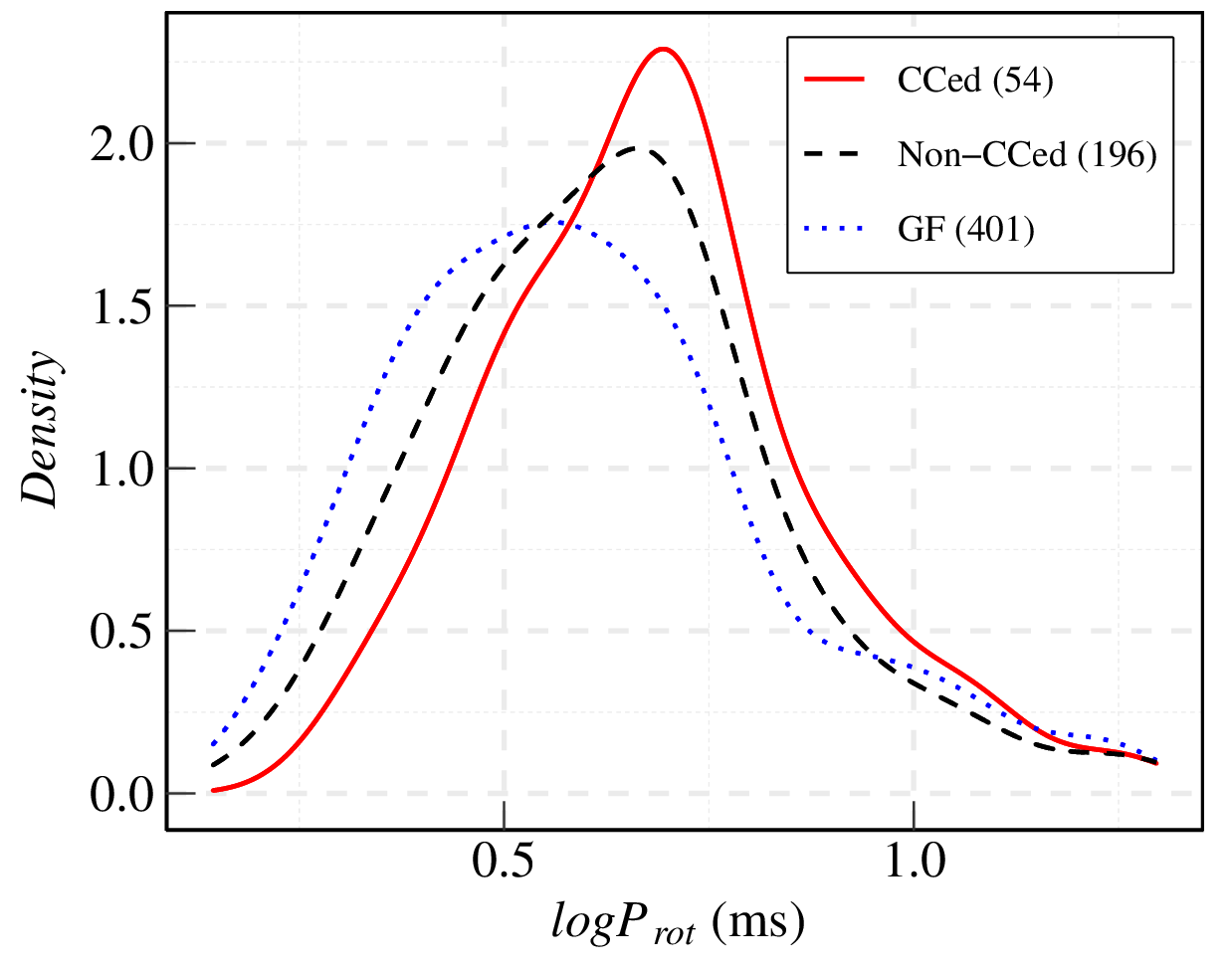}\par
			\includegraphics[width=\linewidth]{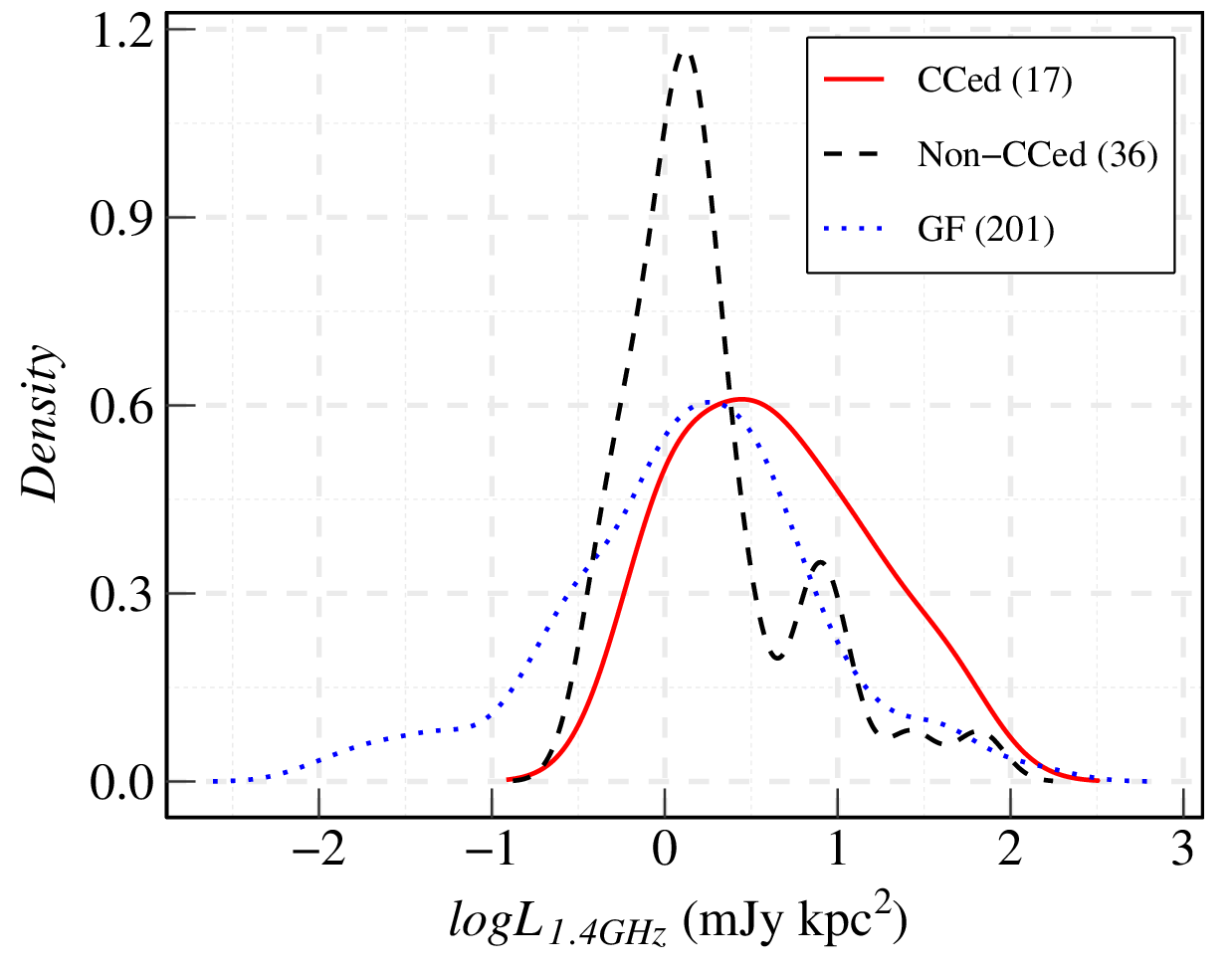}\par
		\end{multicols}
		\vspace{-0.5cm}
		\caption{Kernel density estimates for the distributions of $\log P_{\rm rot}$ and $\log L_{\rm 1.4GHz}$ of MSPs in CCed GCs, Non-CCed GCs, and the GF.}
		\label{fig:kde}    
	\end{figure*}

	Since $P_{\rm rot}$ and $L_{\rm 1.4GHz}$ of the MSPs in CCed GCs are found to be significantly different from those in Non-CCed GCs and the GF, we have further examined their distributions by computing the kernel density estimates (KDEs). The results are shown in \autoref{fig:kde}.  
	In the panel of $P_{\rm rot}$, it clearly shows that the peaks of density distributions systematically shifted towards the larger value from the GF (which lacks dynamical interactions) to the CCed GCs (which have the largest disruption rates among three populations; see \autoref{tab:gcs}). The peaks for the KDEs of $P_{\rm rot}$ for GF, Non-CCed GCs, and CCed GCs are 3.6~ms, 4.6~ms, and 5.0~ms respectively. For $L_{\rm 1.4GHz}$, the KDEs of GF and CCed GC populations peaked at 1.8~mJy~kpc$^{2}$ and 2.8~mJy~kpc$^{2}$ respectively. In the case of Non-CCed GC MSPs, it is interesting to note that there appear to have two peaks in its $L_{\rm 1.4GHz}$ KDE which is located at 1.3~mJy~kpc$^{2}$ and 8.0~mJy~kpc$^{2}$. However, there are only 7 Non-CCed GC MSPs $\gtrsim3$~mJy~kpc$^{2}$ in the current sample which does not allow us to determine whether such multi-modal distributions are genuine or simply a fluctuation due to the small sample. 
	
	\cite{Verbunt2014.561} have compared the fraction of isolated pulsars in GCs with the corresponding disruption rate $\gamma\propto\rho_{c}^{0.5}r_{c}^{-1}$, where $\rho_{c}$ and $r_{c}$ represent the central density and core radius, respectively \citep[cf. Tab. 1 in ][]{Verbunt2014.561}. In their work, they considered a sample of only 14 GCs. Since our sample is now almost three times larger, it is legitimate to revisit this comparison. For computing $\gamma$, we adopted $\rho_{0}$ and $r_{c}$ from \citet[][2010 edition]{Harris1996.112}. In \autoref{tab:gcs}, we compare the numbers of isolated pulsars $N_{s}$ and binary pulsars $N_{b}$ in 37 GCs with their corresponding $\gamma$. GLIMPSE01 is excluded in this part of the analysis because we cannot find its structural parameters in the literature.
	
	We proceeded to examine if there is any correlation between the fraction of isolated pulsars $f_{s}=N_{s}/(N_{s}+N_{b})$ and $\gamma$ with the non-parametric Spearman's rank test, which yields a $p$-value of 0.014. This indicates the correlation between these two quantities is significant. This prompts us to perform a regression analysis to obtain an empirical relation between $f_{s}$ and $\gamma$. However, in view of the small statistics of pulsar population in most GCs, we notice that the $f_{s}$ is very sensitive to $N_{s}$ and $N_{b}$. In particular, many of the GCs have $f_{s}=0$ (\autoref{tab:gcs}). 
	
	To address this issue, we found that Laplace smoothing is a well-established technique in handling categorical data with a small sample size \citep[e.g.][]{manning_2008,gelman_2013}. By adding a smoothing parameter $\alpha$ to the observed counts, the method can stabilize the estimates and avoid zero empirical probabilities. With Laplace smoothing, we obtained the smoothed estimate of $f_{s}$ as $\hat{f}_{s}=\frac{N_{s} + \alpha}{N_{s}+N_{b}+ 2\alpha}$ with $\alpha$ taken to be 1.
	
\begin{figure}
		\includegraphics[width=\columnwidth]{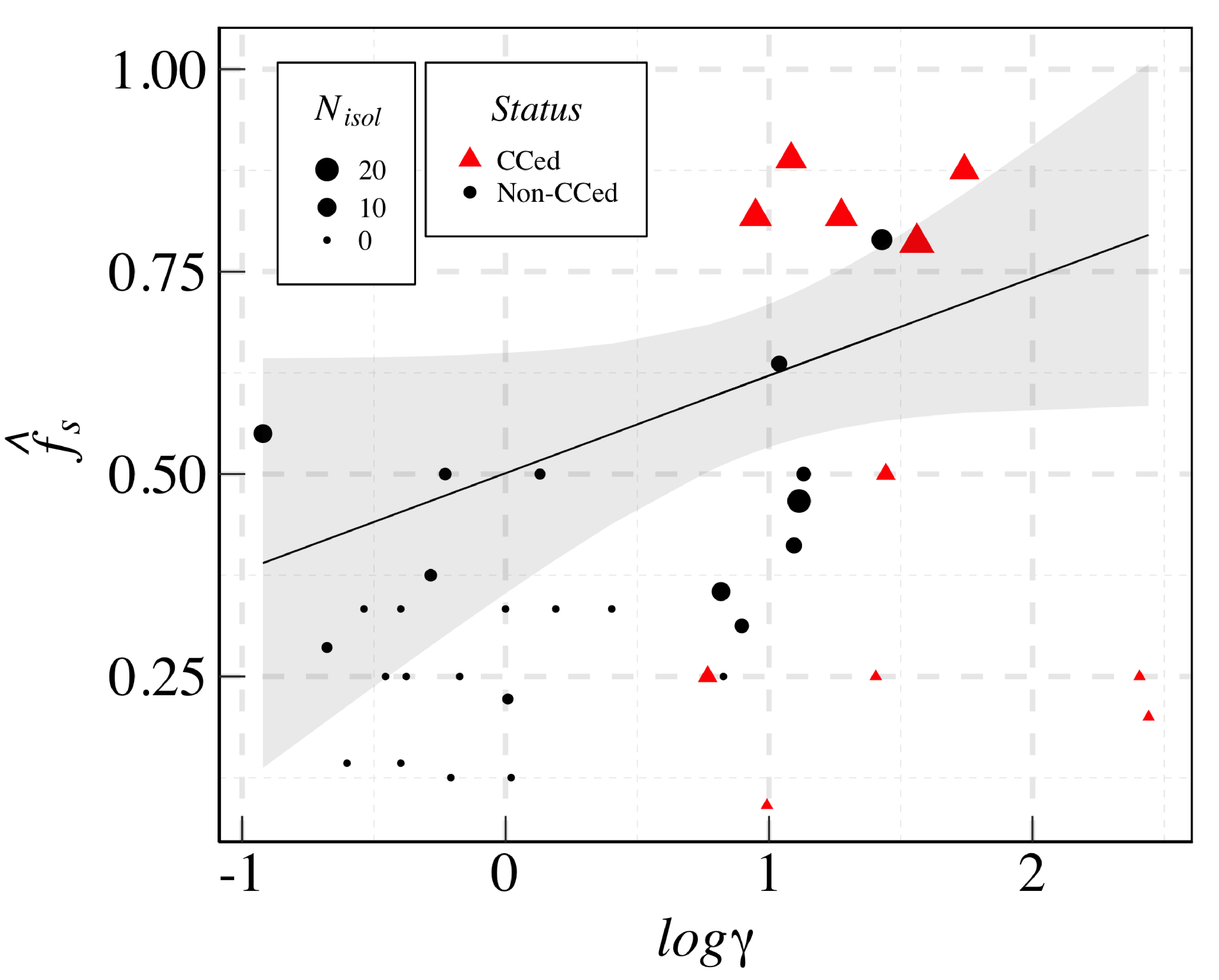}
		\vspace{-0.5cm}
		\caption{Relation between the fraction of isolated pulsar estimated by Laplace smoothing $\hat{f}_{s}$ and the disruption rate $\log\gamma$. The symbol sizes scales with the actual number of observed isolated pulsars. The straight line represents the best-fit linear model with $95\%$ confidence band illustrated by the shaded region.}
		\label{fig:iso_gamma}
	\end{figure}

	In \autoref{fig:iso_gamma}, we show the scatter plot between $\hat{f}_{s}$ and 
	$\log\gamma$ of our sample. It is obvious that the disruption rates of CCed GCs are generally larger than those of Non-CCed GCs. Furthermore, GCs with $\hat{f}_{s}\gtrsim0.5$ are predominantly CCed GCs with $\gamma$ more than ten times larger than the conventional reference level in M4. These findings are fully consistent with the results reported by \cite{Verbunt2014.561}. By fitting a linear model $\hat{f}_{s}=a\log\gamma + b$ to the data with each GC weighted by the numbers of detected pulsars, we found the best-fit parameters of $a=0.12\pm 0.05$ and $b=0.38\pm 0.05$ ($1\sigma$ uncertainties) for this empirical relation.
	To test whether the result of linear regression is sensitive to the adopted smoothing parameter, we repeated the analysis by varying $\alpha$ from 2 to 5. We found that the results obtained from different $\alpha$ values all lie within the 95\% confidence band shown in \autoref{fig:iso_gamma} for the case of $\alpha=1$. 
	
	\begin{table}
		\begin{threeparttable}
			\resizebox{\columnwidth}{!}{
				\begin{tabular}{lc||ccc}
					\toprule
					& {CCed vs non-CCed}\tnote{1} & CCed vs GF\tnote{2} & non-CCed vs GF\tnote{2} & GCs vs GF\tnote{3} \\ 
					\midrule
					$P_{\rm rot}$    & 0.003  & 0.002   & 0.023             & 0.001      \\
					$P_{b}$          & 0.242  & 0.001   & $9\times10^{-5}$  & $10^{-7}$  \\
					$L_{\rm 1.4GHz}$ & 0.014  & 0.001   & 0.094             & 0.041      \\
					$L_{x}$          & 0.315  & 0.137   & 0.078             & 0.030      \\
					\bottomrule
				\end{tabular}
			}
			\caption{Null hypothesis probabilities of A-D test for comparing $P_{\rm rot}$, $P_{b}$, $L_{\rm 1.4GHz}$ and  $L_{x}$ among CCed GCs, Non-CCed GCs and GF.}
			\label{tab:adtest} 
			\begin{tablenotes}
				\item [1] c.f. \autoref{fig:cdf}. 
				\item [2] c.f. \autoref{fig:cdf2}.  
				\item [3] GCs = CCed + Non-CCed
			\end{tablenotes} 
		\end{threeparttable}  
	\end{table}

	\begin{table}\centering
		\begin{threeparttable}[b]
			\begin{tabular}[t]{ccccccc}
				\toprule
				Name & $N_{b}$ & $N_{s}$ & $r_{c}$ & log $\rho_{c}$        & $\gamma$ & Class       \\ 
				&        &         &  (pc)   & ($L_{\odot}$$pc^{-3}$) & ($\gamma_{M4}$) \\
				\midrule
				\multicolumn{7}{ c }{Non-core-collapsed GCs}\\
				\midrule
				47 Tuc   & 19 & 10 & 0.36 & 4.88 & 6.57    & I \\
				M 10     & 2  & 0  & 0.77 & 3.54 & 0.67    & S \\
				M 12     & 2  & 0  & 0.79 & 3.23 & 0.42    & S \\
				M 13     & 4  & 2  & 0.62 & 3.55 & 0.52    & S \\
				M 14     & 5  & 0  & 0.79 & 3.36 & 0.25    & S \\
				M 2      & 6  & 0  & 0.32 & 4.00 & 1.05    & I \\
				M 22     & 2  & 2  & 1.33 & 3.63 & 0.59    & S \\
				M 28     & 10 & 4  & 0.24 & 4.86 & 7.88    & I \\
				M 3      & 6  & 0  & 0.37 & 3.57 & 0.62    & I \\
				M 4      & 1  & 0  & 1.16 & 3.64 & 1.00    & I \\
				M 5      & 6  & 1  & 0.44 & 3.88 & 1.02    & I \\
				M 53     & 4  & 1  & 0.35 & 3.07 & 0.21    & S \\
				M 71     & 5  & 0  & 0.63 & 2.83 & 0.40    & S \\
				M 92     & 1  & 0  & 0.26 & 4.30 & 2.53    & I \\
				NGC 1851 & 9  & 6  & 0.09 & 5.09 & 12.44   & I \\
				NGC 5986 & 1  & 0  & 0.47 & 3.41 & 0.40    & S \\
				NGC 6440 & 4  & 4  & 0.14 & 5.24 & 13.53   & I \\
				NGC 6441 & 3  & 6  & 0.13 & 5.26 & 10.93   & I \\
				NGC 6517 & 3  & 14 & 0.06 & 5.29 & 26.82   & I \\
				NGC 6539 & 1  & 0  & 0.38 & 4.15 & 1.55    & I \\
				NGC 6652 & 2  & 0  & 0.1  & 4.48 & 6.71    & I \\
				NGC 6712 & 1  & 0  & 0.76 & 3.18 & 0.29    & S \\
				NGC 6749 & 2  & 0  & 0.62 & 3.30 & 0.35    & S \\
				NGC 6760 & 1  & 1  & 0.34 & 3.89 & 1.35    & I \\
				Omega Cen   & 8  & 10 & 2.37 & 3.15 & 0.12 & S \\
				Terzan 5 & 24 & 20 & 0.16 & 5.14 & 13.00   & I \\
				\midrule
				\multicolumn{7}{ c }{Core-collapsed GCs} \\
				\midrule
				M 15     & 1  & 8  & 0.14 & 5.05 & 8.89   & D \\
				M 30     & 2  & 0  & 0.06 & 5.01 & 25.42  & D \\
				M 62     & 9  & 0  & 0.22 & 5.16 & 9.82   & I \\
				NGC 362  & 5  & 1  & 0.18 & 4.74 & 5.85   & I \\
				NGC 6342 & 1  & 1  & 0.05 & 4.97 & 27.76  & D \\
				NGC 6397 & 2  & 0  & 0.05 & 5.76 & 254.79 & D \\
				NGC 6522 & 0  & 6  & 0.05 & 5.48 & 55.13  & D \\
				NGC 6544 & 3  & 0  & 0.05 & 6.06 & 275.92 & I \\
				NGC 6624 & 2  & 10 & 0.06 & 5.30 & 36.40  & D \\
				NGC 6752 & 1  & 8  & 0.17 & 5.04 & 18.81  & D \\
				Terzan 1 & 0  & 7  & 0.04 & 3.85 & 12.13  & D \\
				\bottomrule
			\end{tabular}
			\begin{tablenotes}
				\item[*] \textbf{Note} : Number of binary pulsars $N_{b}$ and isolated pulsars $N_{s}$ from \citet{Manchester2005.129}. Core radius $r_{c}$ and central luminosity density, $\rho_{c}$ from \citet[][2010 edition]{Harris1996.112}. Disruption rates $\gamma \propto \rho_{c}^{0.5} r_{c}^{-1}$ from Eq. 2 in \citet[][]{Verbunt2014.561}, which are normalized with the value of M4. The class labels in the seventh column represent the groups of Sparse (S), Intermediate (I), and Dense (D) as determined by GMM (See Sec. 3.2).
			\end{tablenotes}
			\caption{Updated statistics of single and binary pulsars as well as the structural parameters of GCs.}
			\label{tab:gcs}
		\end{threeparttable}
	\end{table}

	\subsection{Alternative Classification by Unsupervised Clustering}
	
	While the aforementioned analyses show that $P_{\rm rot}$ and $L_{\rm 1.4GHz}$ of the MSPs in CCed GCs and Non-CCed GCs are significantly different, the possible ambiguity in the conventional CCed/Non-CCed classification can hamper the robustness of this conclusion. As we have mentioned in Section 2, such classification is determined by the structure of their brightness profiles. In case the central part of GC is poorly resolved, the CCed/Non-CCed classifications are subjected to uncertainties. 
	
	This concern is reflected by the central concentration parameters $c$ given in \citet[][2010 version]{Harris1996.112}, which is defined as the logarithm of the ratio between tidal radius $r_t$ and core radius $r_c$. $c$ is deduced from surface brightness profile fitting \citep{1995AJ....109..218T,1966AJ.....71...64K}. For most of the CCed GCs, no reasonable fit can be obtained and an upper bound of $c=2.5$ is placed instead \citep[cf.][]{1995AJ....109..218T,Harris1996.112}. While $c$ can provide a simple parameter for characterizing the structure, we realize that our sample spans the ranges of $c=0.79-2.07$ and $c=1.63-2.5$ for Non-CCed and CCed GCs, respectively. Such heavily-overlapped ranges of $c$ indicate the CCed/Non-CCed classification in \citet[][2010 version]{Harris1996.112} is not without ambiguity. 
	
	On the other hand, the disruption rates $\gamma$ in \autoref{tab:gcs} might provide a more quantitative measure of the dynamical status of a GC \citep{verbunt2014}. For example, in \autoref{fig:iso_gamma}, we have seen that the fraction of isolated pulsars $f_{s}$ is generally correlated with $\gamma$, though the spread of the data from the best-fit linear model is rather wide. 
	
	Individually, the parameters $\gamma$ and $c$ might not allow an unambiguous classification of GCs. This motivates us to examine whether the classification can be improved by combining both parameters.
	
\begin{figure}
		\includegraphics[width=\columnwidth]{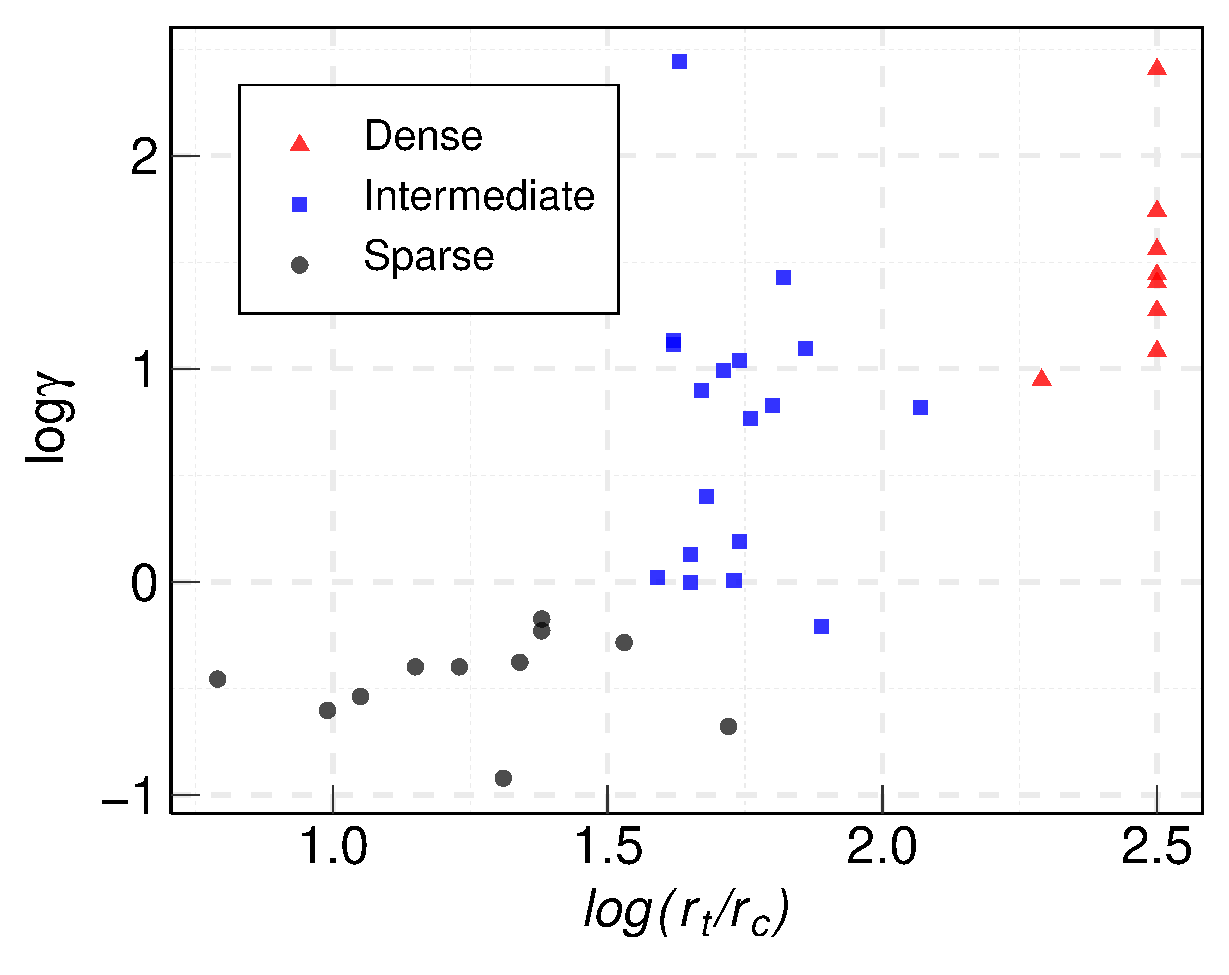}
		\caption{Unsupervised classification of GCs in a plane spanned by the disruption rate $\gamma$ and the central concentration parameter $c$ with the method of 2-dimensional Gaussian Mixture Model (GMM).}
		\label{fig:gmm_cluster}
	\end{figure}

	For deriving the classification rules in the plane spanned by $\gamma$ and $c$, we employed the Gaussian Mixture Model (GMM) algorithm. GMM is a probabilistic model with an assumption that the data originated from a mixture of finite numbers of Gaussian components. We have considered a set of models with the number of mixture components ranging from 1 to 9. We utilized the \texttt{CRAN Mclust} package \citep[version 5.4.6][]{mclust} for the model fitting and computed the likelihoods, $L$, of each model. Model selection is based on the Bayesian information criterion \citep[BIC][]{Schwarz}: ${\rm BIC}= 2\ln{L} -k\ln{N}$, where $k$ and $N$ are the number of estimated parameters and the sample size respectively. We found that the optimal BIC requires three 2-dimensional Gaussian components to model our adopted data in $\gamma-c$ plane. In \autoref{fig:gmm_cluster}, three different groups as clustered by GMM are represented by the symbols of different colour. According to their brightness concentration, we refer to these groups as "Sparse (S)", "Intermediate (I)" and "Dense (D)" hereafter. The corresponding labels of each GC are given in \autoref{tab:gcs}. Under this classification scheme, S group consists of purely Non-CCed GCs and D group only comprises CCed GCs. For the I group, there is a mixture of both Non-CCed and CCed GCs. 
	
\begin{figure*}
		\begin{multicols}{2}
			\includegraphics[width=\linewidth]{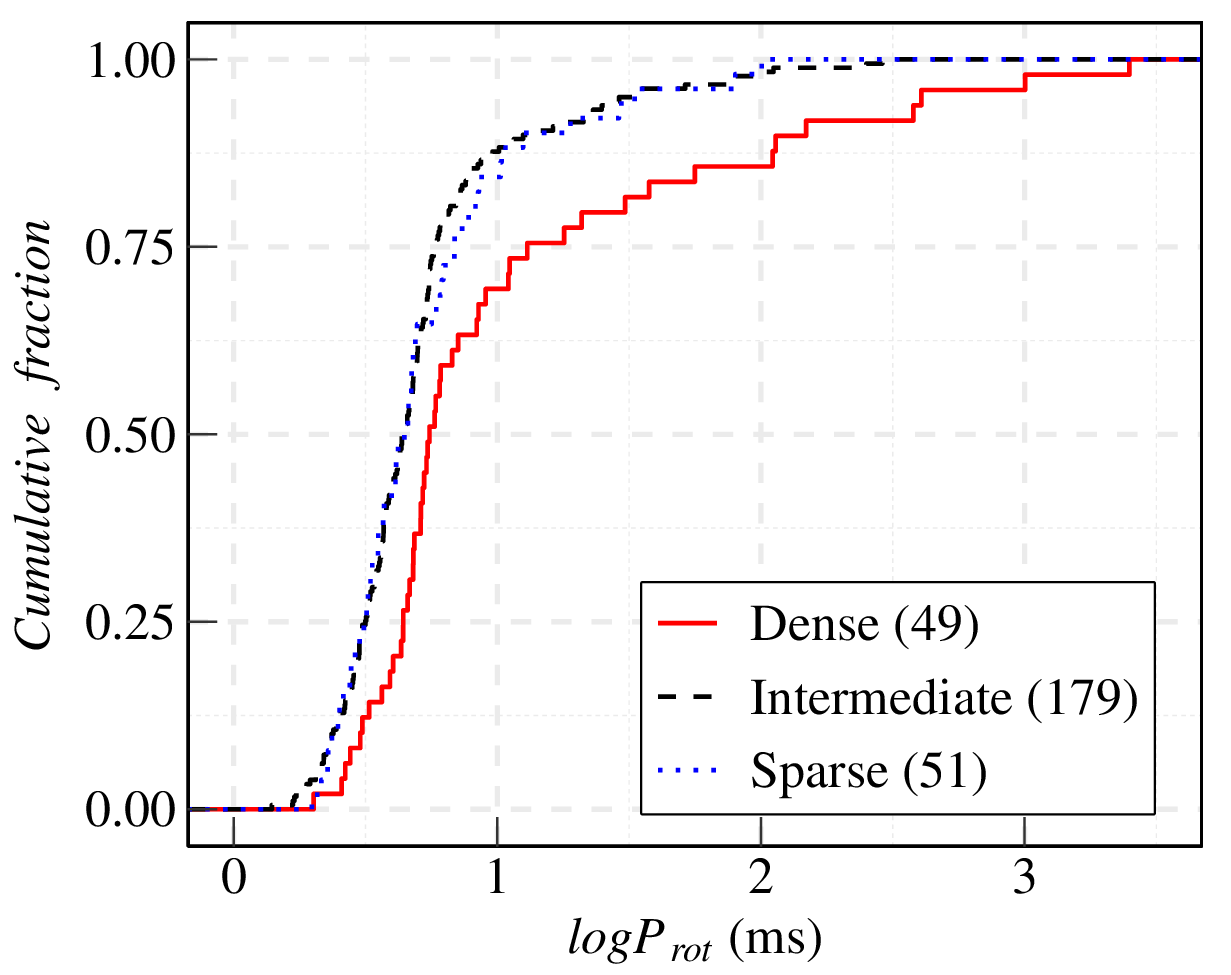}\par
			\includegraphics[width=\linewidth]{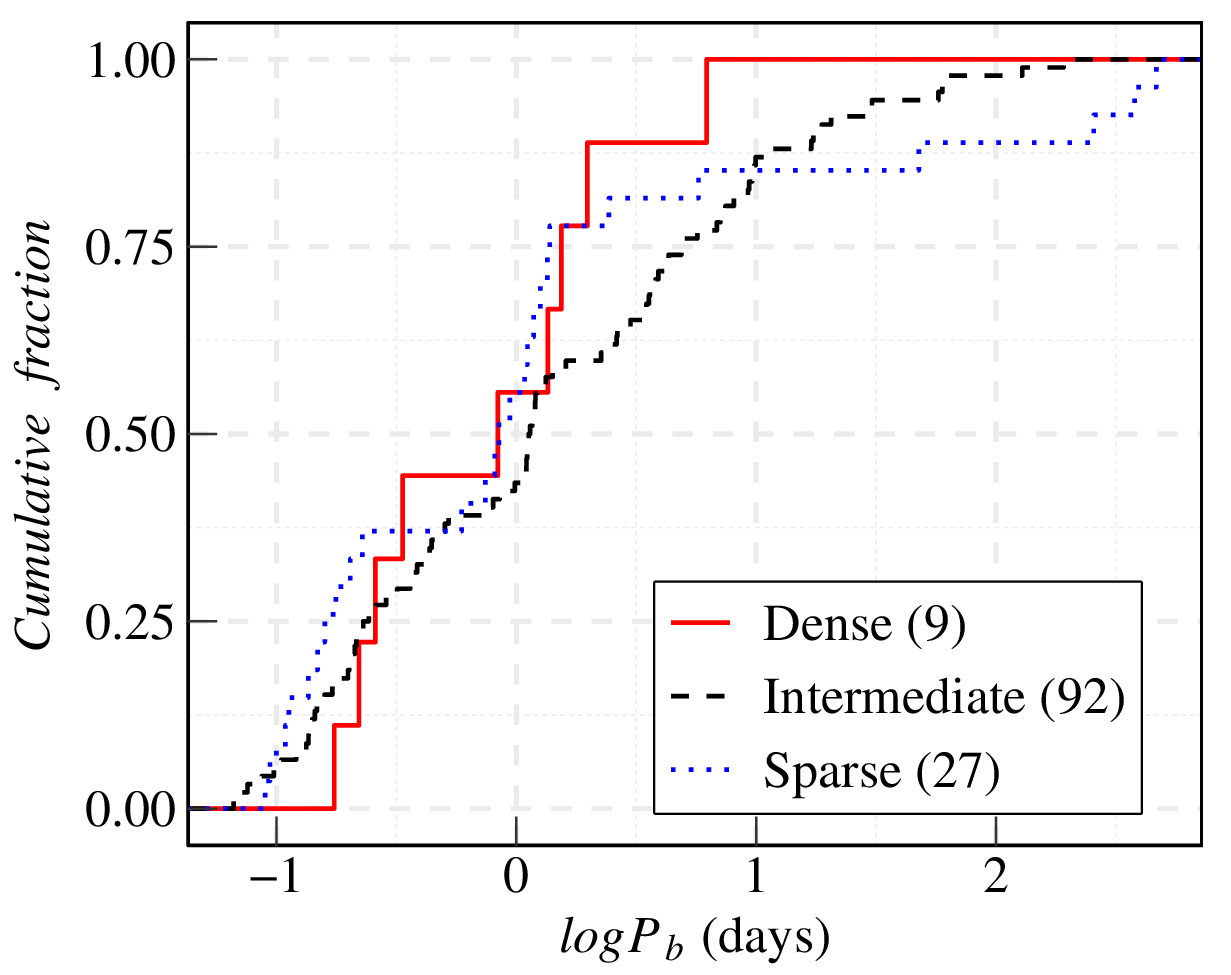}\par
		\end{multicols}
		\begin{multicols}{2}
			\includegraphics[width=\linewidth]{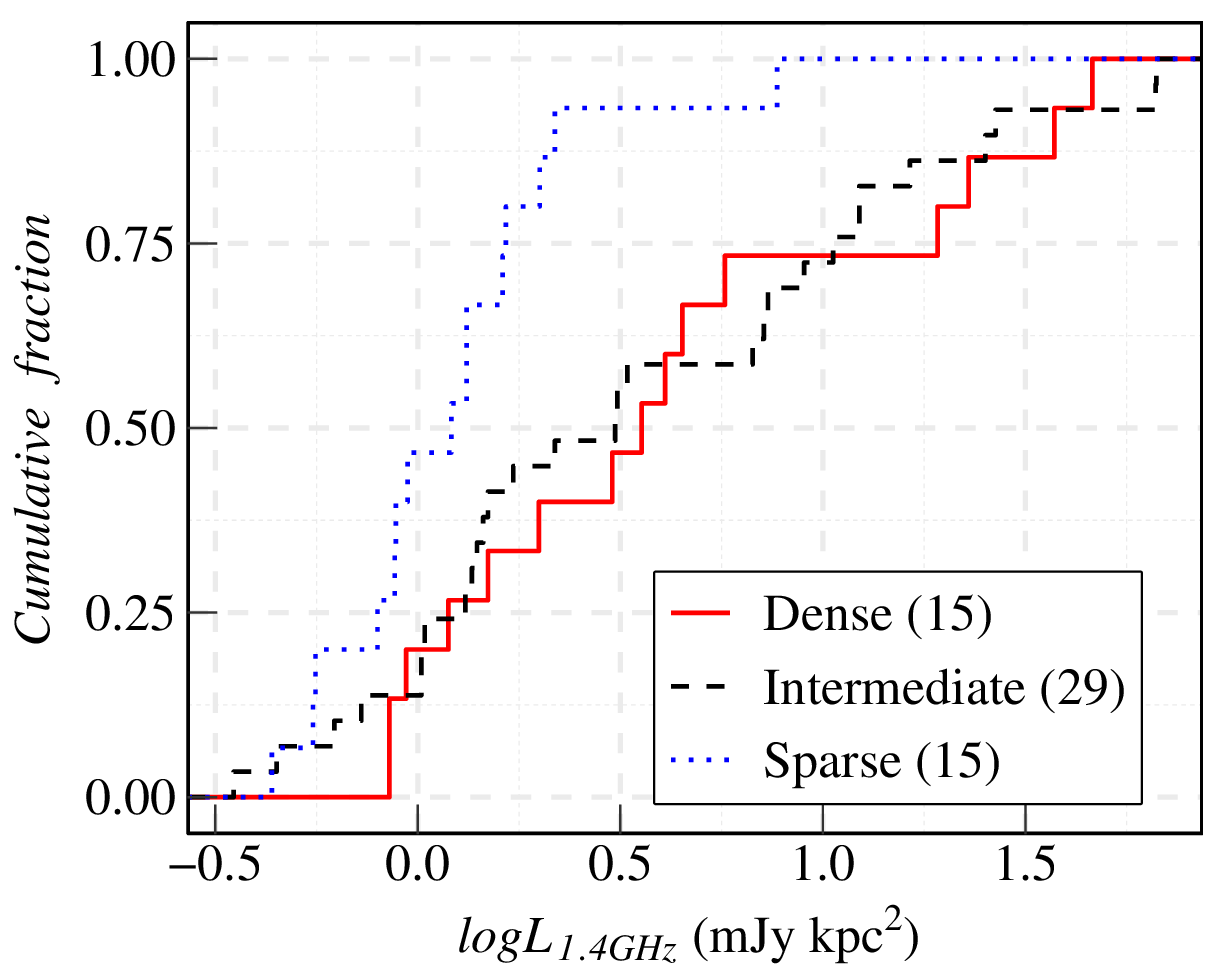}\par
			\includegraphics[width=\linewidth]{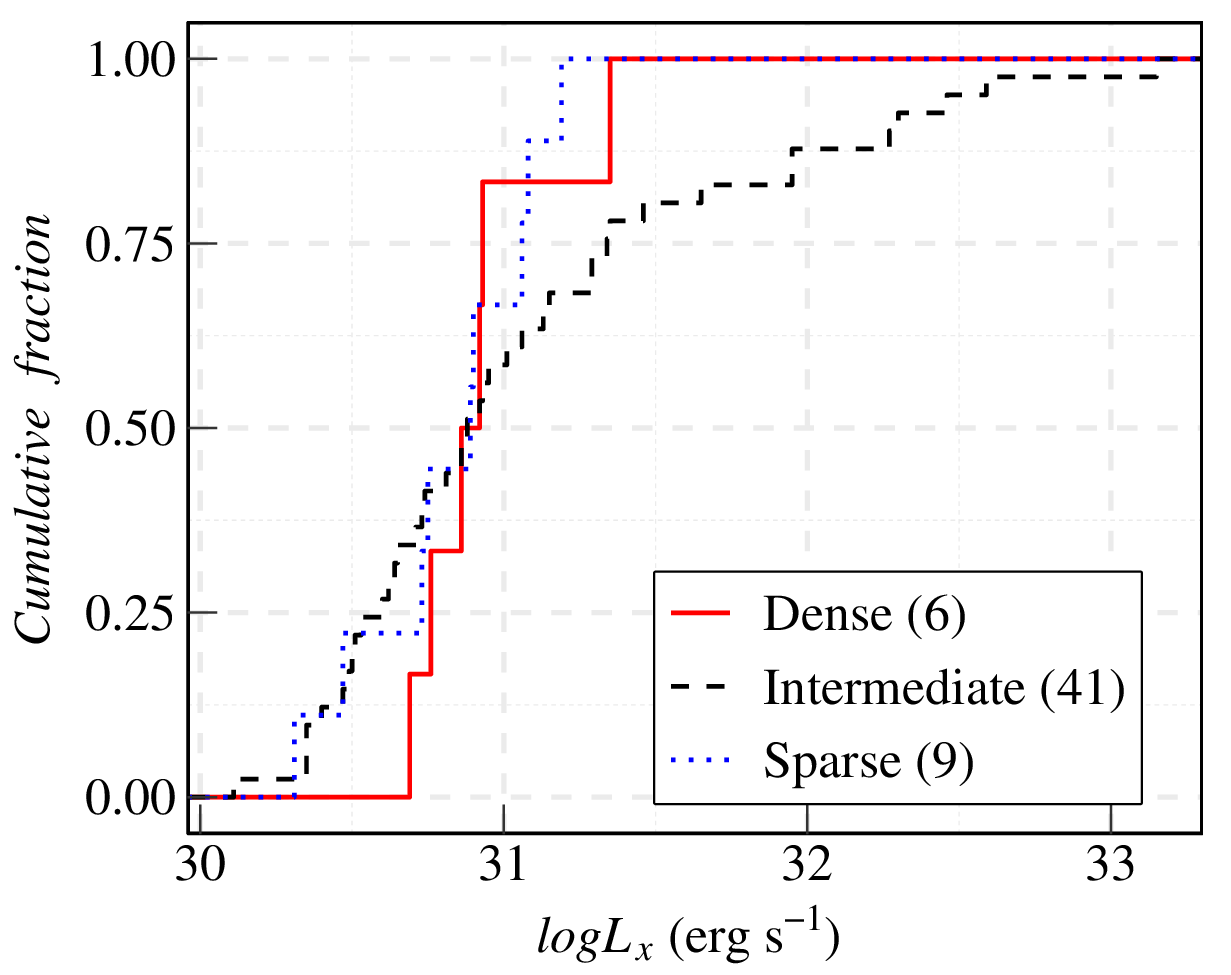}\par
		\end{multicols}
		\vspace{-0.5cm}
		\caption{Comparisons of eCDFs of the selected pulsar properties among S, I and D groups as determined by GMM. The bracketed numbers in the legends show the corresponding sample sizes.}
		\label{fig:cdf_gmm}    
	\end{figure*}

	These three groups in $\gamma-c$ plane are well separated without much overlap (\autoref{fig:gmm_cluster}). The averaged isolated pulsars fractions $\langle f_{s}\rangle$ in S, I and D groups are 0.64, 0.24 and 0.14, respectively, which increase progressively. These suggest such alternative classification is not unreasonable. And this prompts us to re-examine the possible differences of pulsar properties among these three groups. The comparisons of their eCDFs of $P_{\rm rot}$, $P_{b}$, $L_{\rm 1.4GHz}$ and $L_{x}$ are shown in \autoref{fig:cdf_gmm}. The corresponding $p$-values as inferred from the A-D test are summarized in \autoref{tab:gmm_adtest}. 
	
	In comparing $P_{\rm rot}$ between S group and D group, we found that the pulsars in D groups generally rotate slower than those in S group. And such a difference is statistically significant ($p=7\times10^{-3}$). Also, the distribution of $L_{\rm 1.4GHz}$ of S group is found to be significantly different from that of D group  ($p=0.013$) with the pulsars of D group significantly more luminous in L-band than those of S group. 
	These results are fully consistent with those inferred from the comparison between Non-CCed and CCed populations as presented in Section 3.1 (cf. \autoref{fig:cdf} and \autoref{tab:adtest}). 
	
	For I group, which consists of both Non-CCed and CCed GCs, it is obvious that the $P_{\rm rot}$ distribution of I group is very similar to S group (see \autoref{fig:cdf_gmm}). Examining the composition of this group, we found that $\sim90\%$ of the pulsars in I group are originated from Non-CCed GCs which are dominated by the populations in 47~Tuc and Terzan~5. This might apparently account for the similarity.
	Nevertheless, despite the fact that the sample for $L_{\rm 1.4GHz}$ in I group is also dominated by Non-CCed pulsars which have a contribution of 83\%, its distribution is comparable to that of D group. 
	
	We would like to point out that the selection effect on the sample of $L_{\rm 1.4GHz}$ might prevent us from drawing any firm conclusion in comparing this property among these three groups. While the sample size for $P_{\rm rot}$ is 279, there are only 59 pulsars that have their measures of $L_{\rm 1.4GHz}$ available for analysis. This effect is particularly obvious in I group which has its sample size reduced from 179 for $P_{\rm rot}$ to 29 for $L_{\rm 1.4GHz}$.This can be accounted for by the fact that only those sufficiently bright pulsars can have their radio fluxes reliably measured. It is uncertain whether the $L_{\rm 1.4GHz}$ distribution of I group will remain comparable D group when the fainter pulsars are included. Pulsar surveys with improved sensitivity might help to resolve this issue in the future.

	\begin{table}
		\centering
		\begin{tabular}{lccc}
			\toprule
			&  S vs D & S vs I & D vs I \\ 
			\midrule
			$P_{\rm rot}$    & 0.007  & 0.898   & 0.0002    \\
			$P_{b}$          & 0.612  & 0.214   & 0.472    \\
			$L_{\rm 1.4GHz}$ & 0.013  & 0.012   & 0.902    \\
			$L_{x}$          & 0.844  & 0.522   & 0.406    \\
			\bottomrule
		\end{tabular}
		\caption{Null hypothesis probabilities of A-D test for comparing $P_{\rm rot}$, $P_{b}$, $L_{\rm 1.4GHz}$ and  $L_{x}$ among S, I and D groups as classified by GMM.}
		\label{tab:gmm_adtest}  
	\end{table}
	
	\section{Summary \& Discussion}
	Motivated by the recent work by \cite{Lee_2023} which has identified the differences in various properties between the GC and GF pulsar populations, we proceed to investigate whether the variation of intracluster dynamics between CCed and Non-CCed GCs can further diversify the pulsar properties (see \autoref{fig:cdf} and \autoref{fig:cdf2}).

	We found that pulsars in CCed GCs generally rotate slower than those in Non-CCed GCs. This is consistent with the notion that secondary encounters in CCed GCs are enhanced \citep{Verbunt2014.561}, which presumably results in the prevalence of isolated MSPs and fewer X-ray binaries than in Non-CCed GCs with comparable primary encounter rates \citep{Bahramian_2013, Verbunt2014.561, Kremer_2022}. The increased binary disruption efficiency in CCed GCs likely interrupts the angular momentum transfer at an earlier stage of recycling. Consequently, the slower rotation of pulsars in CCed GCs is not unexpected \citep[see also][]{Ivanova_2008}.
	
	\begin{figure}
		\includegraphics[width=\columnwidth]{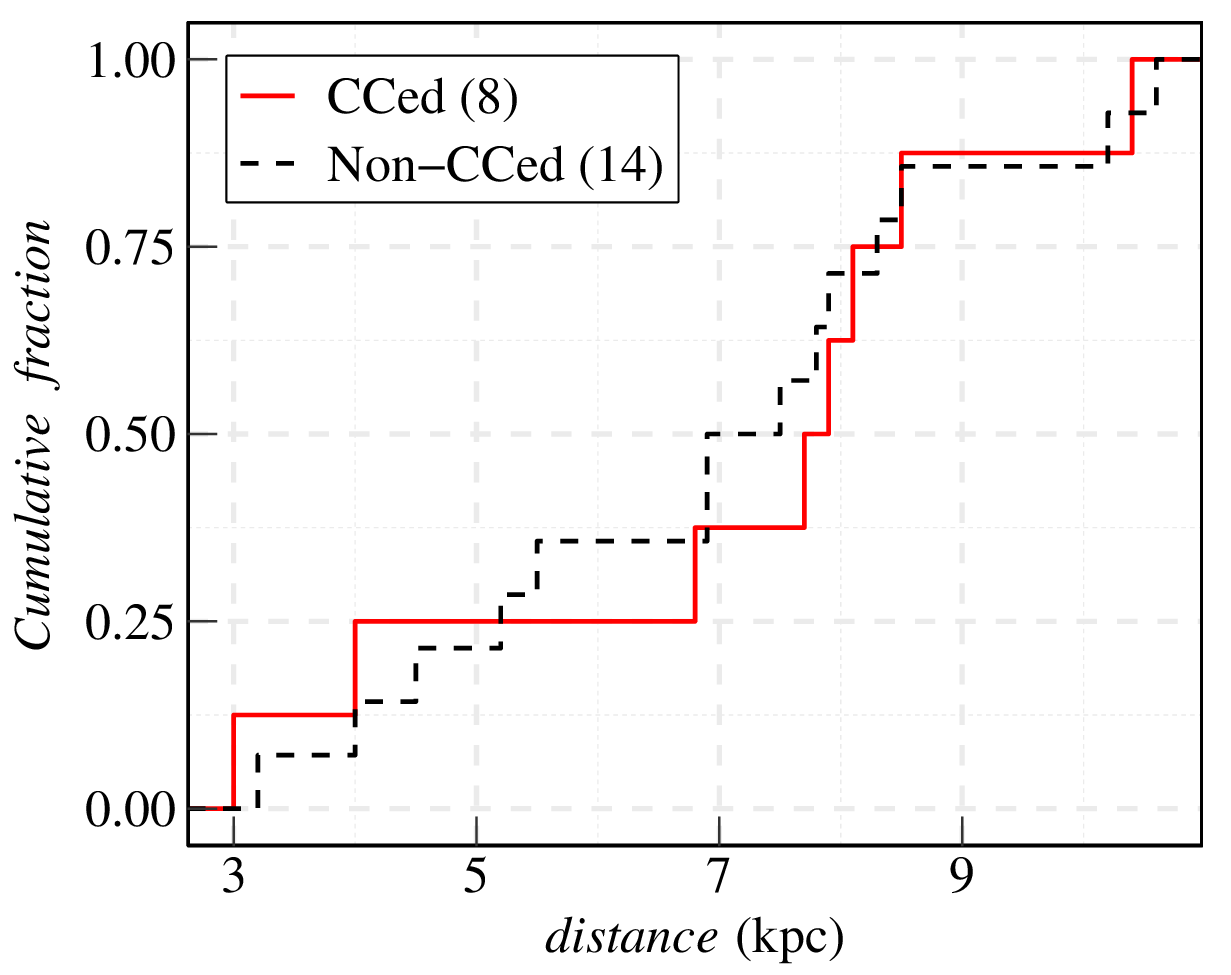}
		\caption{Comparison of eCDFs of the distance between CCed GCs and Non-CCed GCs in our sample.}
		\vspace{-0.5cm}
		\label{fig:dist}
	\end{figure}
	
	\begin{figure}
		\includegraphics[width=\columnwidth]{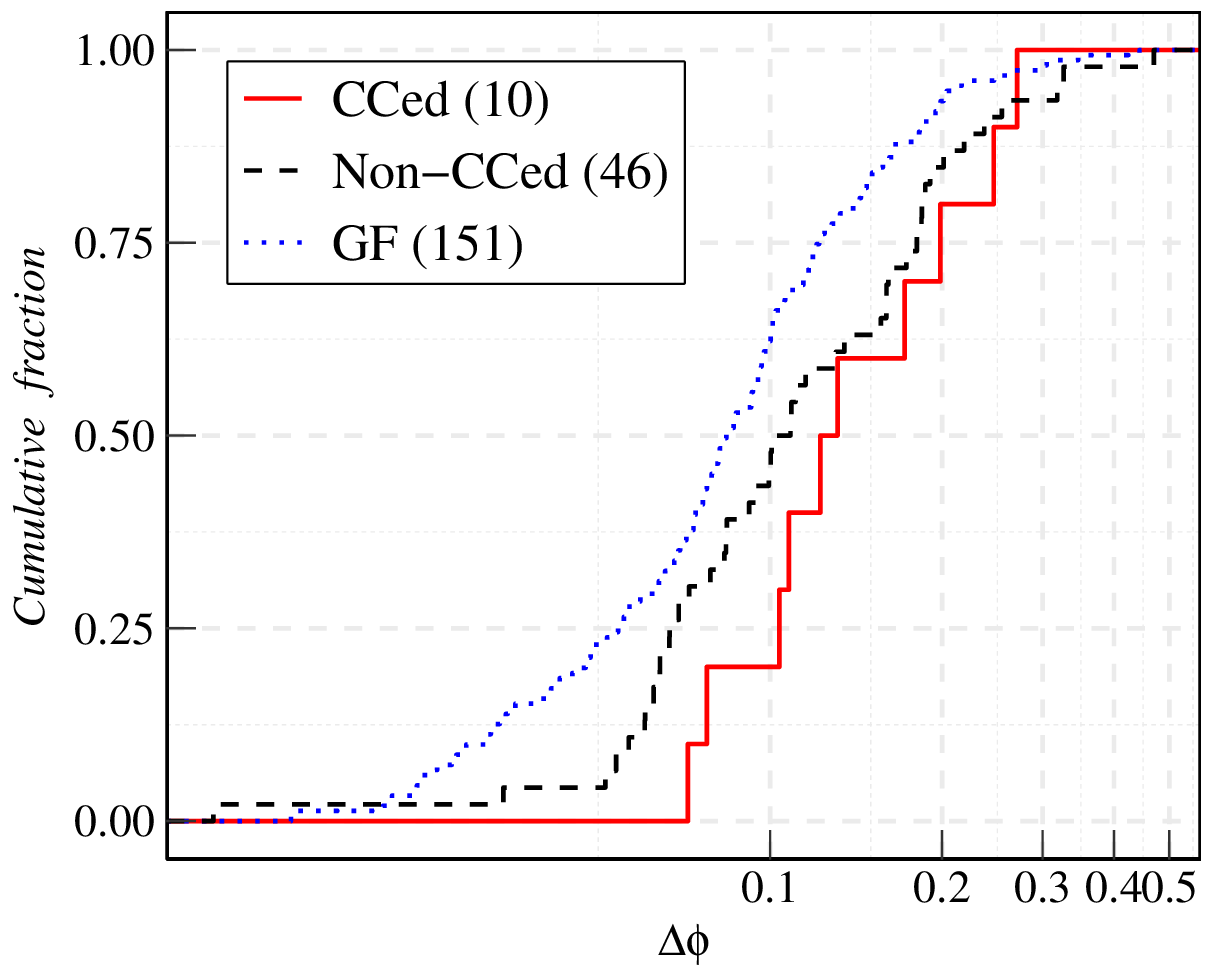}
		\caption{Comparison of eCDFs of the estimates of radio beam sizes $\Delta\phi$ of MSPs in CCed GCs, Non-CCed GCs, and the GF.}
		\vspace{-0.5cm}
		\label{fig:radiobeam}
	\end{figure}

	If the recycling process is halted at an earlier epoch, not only a slower rotating pulsar result, but we should also expect a stronger surface magnetic field than their counterparts in Non-CCed GCs because the magnetic decay due to the mass transfer is suppressed \citep[see the discussion in][]{Lee_2023}. For pulsars, the strength of the dipolar surface magnetic field can be estimated by their rotational period $P_{\rm rot}$ and the corresponding spin-down rate ${\dot{P}_{\rm rot}}$ namely $B_{s}\simeq\sqrt{\frac{3c^{2}I}{2\pi^{2}R_{NS}^{6}}{\dot{P}_{\rm rot}}P_{\rm rot}}$, where $c$ is the speed of light and $R_{NS}$ is the radius of the neutron star. However, such estimation for the pulsars in GCs is complicated by the accelerations in the gravitational potential of a GC, which can bias the measurement of ${\dot{P}_{\rm rot}}$. Up to now, there are only a handful of GC pulsars with their intrinsic ${\dot{P}_{\rm rot}}$ estimated \citep[cf. Tab.~4 in][]{Lee_2023} and therefore, we are not able to directly compare the $B_{s}$ of the pulsars in CCed and Non-CCed GCs. 
	
	On the other hand, as a pulsar radiates by tapping its rotational energy, the radiation power should be proportional to the spin-down power $\dot{E}$ which is expressed as $\dot{E}=4\pi^{2}I{\dot{P}_{\rm rot}}P_{\rm rot}^{-3}\propto B_{s}^{2}P_{\rm rot}^{-4}$ where $I$ is the moment of inertia.  Therefore, the radio luminosity $L_{\rm 1.4GHz}$ can be treated as a proxy for probing $B_{s}$ of the GC pulsars. 
	
	Our analysis indicates that $L_{\rm 1.4GHz}$ of the pulsars in CCed GCs are significantly higher than those in Non-CCed GCs (cf. \autoref{fig:cdf}).  Together with the fact that $P_{\rm rot}$ of CCed GC pulsars are longer than those in Non-CCed GCs, we can infer that $B_{s}$ of CCed GC pulsars are stronger than those in Non-CCed GCs which is in line with our aforementioned speculation.
	
	To investigate whether the difference in $L_{\rm 1.4GHz}$ is genuine, we have further checked whether such a difference can be a result of the observational effect. If a GC is close to us, a flux-limited survey will uncover more faint sources than those in the more distant GCs. For examining this issue, we compared the distances $d$ between the CCed and Non-CCed GCs in our sample, and the results are shown in \autoref{fig:dist}. The medians of $d$ of CCed and Non-CCed GCs are 6.8~kpc and 6.9~kpc respectively. With A-D test, we do not find any significant difference between these two eCDFs ($p>0.05$). And hence, we conclude that the difference in $L_{\rm 1.4GHz}$ between CCed and Non-CCed GCs is genuine.
	
	On the other hand, the A-D test indicates that the differences in $L_{\rm 1.4GHz}$ and $L_{x}$ between the GF and GC MSP populations are statistically significant. However, we notice that many GF MSPs are located in our proximity. The medians of $d$ for radio-selected MSPs in GCs and GF in our sample are found to be 6.9 kpc and 1.7 kpc, respectively. A-D test yields a $p$-value of $\sim10^{-22}$ which indicates a very significant difference between their distributions of $d$. Consequently, the excess at the lower end of the distribution of $L_{\rm 1.4GHz}$ for GF MSPs (\autoref{fig:cdf2}) can be a result of observational bias. This bias also affects the comparison of $L_{x}$ between MSPs in GCs (median $d=4.9$ kpc) and GF (median $d=1.2$ kpc) in our sample.
	
	In conclusion, our results demonstrate that CCed and Non-CCed GC pulsar populations exhibit differences in their rotation rates and radio luminosities, with CCed GC pulsars rotating slower and having higher radio luminosities. This supports the idea that the recycling process is halted earlier in CCed GCs, leading to stronger surface magnetic fields and slower rotations. 
	
	For further examining the effect of dynamical effects on the structure of the surface magnetic field, we would like to compare the radio beam sizes of MSPs in GF, CCed GCs, and Non-CCed GCs. The beam sizes can be estimated by $\Delta\phi=W_{50}/P_{\rm rot}$, where $W_{50}$ is the pulse width at 50\% of the peak in the unit of time as obtained from the ATNF catalog \citep{Manchester2005.129}. The comparisons of $\Delta\phi$ among three populations are given in \autoref{fig:radiobeam}. 
	
	It is interesting to note that the $\Delta\phi$ of MSPs in GF is smaller than those in Non-CCed and CCed GCs. With A-D test, we find $\Delta\phi$ of the GF population is significantly smaller than those of Non-CCed MSPs ($p$-value$\sim$0.01) and CCed MSPs ($p$-value$\sim$0.02). We also note that the $\Delta\phi$ from Non-CCed GCs is apparently smaller than that from CCed GCs, although the A-D test does not yield a $p$-value below our pre-defined criterion.
	
	These results conform with the expectation that different recycling histories can lead to different surface magnetic field structures. \cite{chen_ruderman_1993} argued that mass accretion could reduce the polar cap radius and hence the size of the open field line region. This notion is supported by \cite{kramer_1998}, which found the open angle of GF MSPs is smaller than that expected from the dipolar geometry (cf. Fig. 12 in their paper). 
	
	The fact that the $\Delta\phi$ of GF MSPs is smaller than those of GC MSPs is consistent with the scenario that the accretion phase of GC MSPs is shortened by dynamical disruption, as suggested by \cite{Lee_2023}. Since the disruption rate is generally higher in CCed GCs (see \autoref{tab:gcs} \& \autoref{fig:iso_gamma}), the MSPs in CCed GCs should have a larger beam size than those in Non-CCed GCs. However, a firm conclusion is precluded by the current sample size. With more samples available in the future, the comparison of $\Delta\phi$ between these two classes of GC MSPs should be revisited. 
	
	We have to point out a caveat in the comparison of $\Delta\phi$ presented here. First, owing to the complexity of the radio pulse profile, $W_{50}$ should be considered as a poor estimator for the size of the emission beam. Second, beam size should be a function of observing frequency. However, such information is not available in the ATNF catalog. A more accurate determination of the emission geometry should be derived from fitting the polarization data. Therefore, we strongly encourage a dedicated study to compare the emission geometry of MSPs in GF and GCs with radio polarization, which can help to scrutinize our hypothesis.
	
	Lastly, we would like to emphasize that all the aforementioned discussions are based on the conventional CCed/Non-CCed classification of GCs, which relies on photometric measurements \citep{1995AJ....109..218T,Harris1996.112}. In Section 3.2, we have pointed out a possible ambiguity of this conventional classification scheme. \cite{Bianchini2018} have also mentioned that there is no robust connection between the photometric central concentration and the dynamical state of a GC. 
	
	By combining the central concentration parameter $c$ and a dynamical measure of disruption rate $\gamma$, we have shown that the GCs in our sample can be divided into three groups (\autoref{fig:gmm_cluster}). For two groups maximally separated in the $\gamma-c$ plane, namely S group and D group, they purely comprised Non-CCed GCs and CCed GCs, respectively (cf. \autoref{tab:gcs}). By comparing the distributions of $P_{\rm rot}$ and $L_{\rm 1.4GHz}$ between these two groups, the differences remain to be statistically significant. On the other hand, the intermediate I group has a mixture of CCed and Non-CCed GCs. Both flux-limited samples and a strong bias in I group by the pulsars from a few Non-CCed GCs (e.g. 47~Tuc and Terzan~5) preclude any conclusive comparison with the other two groups.
	
	This has also raised a concern that the classification scheme of GCs might not be unique. In view of the complex evolution of GCs \citep[e.g.][]{Ivanova_2006, Hong_2017}, the description of both dynamical status and structure of GCs can be more complicated than the binary classification as simple as CCed or Non-CCed. For example, by examining the radial distribution of blue stragglers, \cite{Ferraro_2012} have shown that the dynamical age of GCs can be divided into three groups. With a more comprehensive classification scheme proposed by further studies, the differences in pulsar properties among different groupings can be reexamined.

	\vspace{-0.7cm}
	\section*{Acknowledgements}
	K.O is supported by the National Research Foundation of Korea grant 2022R1F1A1073952 and 2022R1A6A3A13071461. C.Y.H. is supported by the research fund of Chungnam National University and by the National Research Foundation of Korea grant 2022R1F1A1073952. J.T. is supported by the National Key Research and Development Program of China (grant No. 2020YFC2201400) and the National Natural Science Foundation of China (NSFC, grant No. 12173014). A.K.H.K. is supported by the National Science and Technology Council of Taiwan through grant 111-2112-M-007-020. P.H.T. is supported by NSFC grant No. 12273122 and the China Manned Space Project (No. CMS-CSST-2021-B09). K.L.L. is supported by the National Science and Technology Council of the Republic of China (Taiwan) through grant 111-2636-M-006-024, and he is also a Yushan Young Fellow supported by the Ministry of Education of the Republic of China (Taiwan).
	
	
	\section*{Data Availability}
	The data underlying this article were accessed from Chandra Data Archive (https://cda.harvard.edu/chaser/) and ATNF (https://www.atnf.csiro.au/research/pulsar/psrcat/).
	
	
	\vspace{-0.7cm}
	\bibliographystyle{mnras}
	\bibliography{reference.bib}
	
	
	\bsp	
	\label{lastpage}
\end{document}